\documentclass[12pt]{article}
\usepackage[dvips]{graphics}

\begin{document}
\begin{center}
\LARGE {The standing wave model of \\the mesons and baryons}
\bigskip

\Large {E.L. Koschmieder}
\medskip

\small {Center for Statistical Mechanics\\
The University of Texas at Austin, Austin TX 78712\\
e-mail: koschmieder@mail.utexas.edu}
\bigskip

\Large  {November 24, 2002}
\end{center}
\bigskip

\noindent
\small
{Only photons are needed to explain the masses of the 
$\pi^0$,\,$\eta$,\,$\Lambda$,\,$\Sigma^0$,\,$\Xi^0$,\,$\Omega^-$,\,$\Lambda_c^+$,\,$\Sigma_c^0$,\\$\Xi_c^0$ 
and $\Omega_c^0$ mesons and baryons with the sum of the energies contained in the 
frequencies of standing electromagnetic waves in a cubic black body. Only 
neutrinos are needed to explain the mass of the $\pi^\pm$ mesons with the sum 
of the energies of standing oscillations of muon and electron neutrinos in 
a cubic lattice plus the energies contained in the rest masses of the 
neutrinos. Neutrinos and photons are needed to explain the masses of the 
K-mesons, the neutron and the D-mesons. Surprisingly the mass of the $\mu^\pm$ mesons can also 
be explained without an additional assumption by the oscillation energies 
and rest masses of a neutrino lattice. From the difference of the masses of the $\pi^\pm$ mesons 
and $\mu^\pm$ mesons we find that the rest mass of the muon neutrino is 
47.5\,milli-eV/c$^2$. From 
the difference of the masses of the neutron and proton we find that the rest 
mass of the electron neutrino is 
 0.55\,meV/c$^2$.   The potential of the weak force between the lattice points can be determined 
from Born's lattice theory. From the weak force between the lattice points follows 
automatically the existence of a strong force between the sides of two 
lattices. The strong nuclear force is the sum of the unsaturated weak forces at 
the sides of each lattice and is therefore about 10$^6$ times stronger 
than the weak force.}

\normalsize

\section*{Introduction}

A theory giving an equation 
whose eigenvalues would determine the masses of the elementary particles 
has sofar not been found. The so-called ``Standard Model" of the particles 
has, until now, not come up with a precise determination of the masses of 
either the mesons and baryons or of the leptons, which means that neither 
the mass of the fundamental electron nor the mass of the fundamental 
proton have been explained. This is so although the quarks, the foundation 
of the standard model, have been introduced by Gel-Mann [1] nearly forty 
years ago. The difficulties explaining the masses of the mesons and 
baryons with the standard model seem to be related to the uncertainty of 
the masses of the quarks for which values are considered ranging from zero 
rest mass to values from 2 to 8 MeV for e.g. the u-quark, according to the 
Particle Physics Summary [2], to values on the order of 100 MeV. Suppose one 
has agreed on definite values of the masses of the various quarks then one 
stands before the same problem one has faced with the conventional 
elementary particles, namely one has to explain why the quarks have their 
particular masses and what they are made of. The other most frequently 
referred to theory dealing with the elementary particles, the ``String 
Theory" introduced by Witten [3] about twenty years ago, or its successor 
the superstring theory, have despite their mathematical elegance not led 
to experimentally verifiable results. There are very many other attempts 
to explain the elementary particles or only one of the particles, too many 
to list them here. It appears that there is room for another attempt to 
explain the masses of the elementary particles.

\section {The spectrum of the masses of the particles}

As we have done before [4] we will focus attention on the so-called ``stable" mesons 
and baryons whose masses are reproduced with other data in Tables 1 and 4.
It is obvious that any attempt to explain the masses of the mesons and 
baryons should begin with the particles that are affected by the fewest 
parameters. These are certainly the particles without isospin (I = 0) and 
without spin (J = 0), but also with strangeness S = 0, and charm C = 0. 
Looking at the particles with I,J,S,C = 0 it is startling to find that 
their masses are quite close to integer multiples of the mass of the 
$\pi^0$ meson. It is m($\eta) = (1.0140 \pm 0.0003)\,\cdot$\,4m($\pi^0$), 
and 
the mass of the resonance $\eta^\prime$ is m($\eta^\prime$) = (1.0137 $\pm$ 
0.00015)\,$\cdot$\,7m($\pi^0$). We also note that the average mass ratios 
[m($\eta$)/m($\pi^0$) + m($\eta$)/m($\pi^\pm$)]/2 = 3.9892 = 
0.9973$\,\cdot$\,4 and [m($\eta^\prime$)/m($\pi^0$) + 
m($\eta^\prime$)/m($\pi^\pm$)]/2 = 6.9791 = 0.9970\,$\cdot$\,7 are good 
approximations to the integers 4 and 7. Three particles seem hardly to be 
sufficient to establish a rule. However, if we look a little further we 
find that m($\Lambda$) = 1.0332\,$\cdot$\,8m($\pi^0$) or m($\Lambda$) = 
1.0190\,$\cdot$\,2m($\eta$). We note that the $\Lambda$ particle has spin 
1/2, 
not spin 0 as the $\pi^0$,\,$\eta$ mesons. Nevertheless, the mass of 
$\Lambda$ is close to 8m($\pi^0$). Furthermore we have m($\Sigma^0$) = 
0.9817\,$\cdot\,$9m($\pi^0$), m($\Xi^0) = 0.9742\,\cdot\,$10m($\pi^0$), 
m$(\Omega^-)$ = 
1.0325\,$\cdot\,$12m($\pi^0)$ = 1.0183\,$\cdot$\,3m($\eta$), ($\Omega^-$ is 
charged 
and has spin 3/2). Finally the masses of the charmed baryons are 
m($\Lambda_c^+$) = 0.9958\,$\cdot$\,17m($\pi^0$) = 
1.024\,$\cdot$\,2m($\Lambda$), m($\Sigma_c^0$) = 
1.0093\,$\cdot$\,18m($\pi^0$), 
m($\Xi_c^0$) = 1.0167\,$\cdot$\,18m($\pi^0$), and m($\Omega_c^0$) = 
1.0017\,$\cdot$\,20m($\pi^0$).
 
	\begin{table}\caption{The $\gamma$-branch of the particle 
spectrum}
	\begin{tabular}{lllllcl}\\
\hline\hline\\
 & m/m($\pi^0$) & multiples & decays & fraction & spin & mode 
 \\
 & & & & (\%) & & \\
[0.5ex]\hline
\\
$\pi^0$ & 1.0000 & 1.0000\,\,$\cdot$\,\,$\pi^0$ & $\gamma\gamma$ & 98.798 & 0 & (1.1)\\
 & & & $e^+e^-\gamma$ & \,\,\,1.198 & &\\
\\
$\eta$ & 4.0559 & 1.0140\,$\cdot$\,\,4$\pi^0$ & $\gamma\gamma$ & 39.25 & 0 & 
(2.2)\\
 & & & 3$\pi^0$ & 32.1 & &\\
 & & & $\pi^+\pi^-\pi^0$ & 23.2 & &\\
 & & & $\pi^+\pi^-\gamma$ & \,\,\,4.78 & &\\
\\
$\Lambda$ & 8.26577 & 1.0332\,$\cdot$\,\,8$\pi^0$ & p$\pi^-$ & 63.9 & $\frac{1}{2}$ 
& 2$\cdot$(2.2)\\
 & & 1.0190\,$\cdot$\,\,2$\eta$ & n$\pi^0$ & 35.8 & &\\
\\
$\Sigma^0$ & 8.8352 & 0.9817\,$\cdot$\,\,9$\pi^0$ & $\Lambda \gamma$ & 100 & 
$\frac{1}{2}$ & 2$\cdot (2.2) + (1.1)$\\
\\
$\Xi^0$ & 9.7417 & 0.9742\,$\cdot$\,10$\pi^0$ & $\Lambda\pi^0$ & 99.54 & $\frac{1}{2}$ & 
2$\cdot(2.2) + 2(1.1)$\\
\\
$\Omega^-$ & 12.390 & 1.0326\,$\cdot$\,12$\pi^0$ & $\Lambda K^-$ & 67.8 & 
$\frac{3}{2}$ & 3$\cdot(2.2)$\\
 & & 0.9986\,$\cdot$\,12$\pi^\pm$ & $\Xi^0\pi^-$ & 23.6 & &\\
 & & 1.0183\,$\cdot$\,\,3$\eta$ & $\Xi^-\pi^0$ & \,\,\,8.6 & &\\
\\
$\Lambda_c^+$ & 16.928 & 0.9958\,$\cdot$\,17$\pi^0$ & many & & $\frac{1}{2}$ & 
2$\cdot(2.2) + (3.3)$\\
 & & 0.9630\,$\cdot$\,17$\pi^\pm$\\
\\
$\Sigma_c^0$ & 18.167 & 1.0093\,$\cdot$\,18$\pi^0$ & $\Lambda_c^+\pi^-$ & 
$\approx$100 & $\frac{1}{2}$ & $\Lambda_c^+ + (1.1)$\\
\\
$\Xi_c^0$ & 18.302 & 1.0167\,$\cdot$\,18$\pi^0$ & seven & (seen) & $\frac{1}{2}$   & 
2$\cdot(3.3)$\\
\\
$\Omega_c^0$ & 20.033 & 1.0017\,$\cdot$\,20$\pi^0$ & four & (seen) & $\frac{1}{2}$ & 
2$\cdot(3.3) + 2(1.1)$\\
[0.2cm]\hline\hline
	\end{tabular}
	\end{table}

   Now we have enough material to 
formulate the $\emph{integer multiple}$ $\emph{rule}$, according to which the masses of the 
$\eta$,\,$\Lambda$,\,$\Sigma^0$,\,$\Xi^0$,\,$\Omega^-$,\,$\Lambda_c^+$,\,$\Sigma_c^0$,\,$\Xi_c^0$, 
and $\Omega_c^0$ particles are, in a first approximation, integer 
multiples of the mass of the $\pi^0$ meson, although some of the particles 
have spin, and may also have charge as well as strangeness and charm. A 
consequence of the integer multiple rule must be that the ratio of the 
mass of any meson or baryon listed above divided by the mass of another 
meson or baryon listed above is equal to the ratio of two integer numbers. 
And indeed, for example m($\eta$)/m($\pi^0$) is practically two times 
(exactly 0.9950$\,\cdot$\,2) the ratio m($\Lambda$)/m($\eta$). There is 
also 
the ratio m($\Omega^-$)/m($\Lambda$) = 0.9993\,$\cdot$\,3/2. We have 
furthermore e.g. the ratios m($\Lambda$)/m($\eta$) = 1.019\,$\cdot$\,2, 
m($\Omega^-$)/m($\eta$) = 1.018\,$\cdot$\,3, m($\Lambda_c^+$)/m($\Lambda$) 
= 
1.02399\,$\cdot$\,2, m($\Sigma_c^0$)/m($\Sigma^0$) = 1.0281\,$\cdot$\,2, 
and 
m($\Omega_c^0$)/m($\Xi^0$) = 1.0282\,$\cdot$\,2.

   We will call, for reasons to be explained later, the particles 
discussed above, which follow in a first approximation the integer 
multiple rule, the \emph{$\gamma$-branch} of the particle spectrum. The mass 
ratios of these particles are in Table 1. The deviation of the mass ratios 
from exact integer multiples of m($\pi^0$) is at most 3.3\%, the average 
of the factors before the integer multiples of m($\pi^0$) of the nine 
$\gamma$-branch particles in Table 1 is 1.0066 $\pm$ 0.0184. From a least 
square analysis follows that the masses of the ten particles on Table 1 lie on 
a straight line given by the formula

\begin{equation} \mathrm{m}(N)/\mathrm{m}(\pi^0) = 1.0065\,N - 0.0043 \qquad  
N\,\ge\,1, 
\end{equation}

\noindent     
where N is the integer number nearest to the actual ratio of the particle 
mass divided by m($\pi^0$). The correlation coefficient in equation (1) 
has the nearly perfect value r$^2$ = 0.999.
 
   The integer multiple rule applies to more than just the stable mesons 
and baryons. The integer multiple rule applies also to the $\gamma$-branch 
baryon resonances which are listed in Table 2 and the meson resonances in 
Table 3.

	\begin{table}[h]\caption{The $\gamma$-branch baryons with 
I\,$\leq$\,1,\,J\,=\,$\frac{1}{2}$ }  
	\begin{tabular}{lllr||cllr@{,}lr} \\
\hline\hline\\
particle & m/m($\pi^0$) & I,\,\,J  & N & & particle & m/m($\pi^0$) & 
\multicolumn{2}{c}{I,\,\,J} & 
N\\ 
[0.5ex]\hline
\\[-.3cm]
$\Lambda$ & 8.26577 & 0,\,$\frac{1}{2}$ & 8 & &  $\Xi^0$ & 9.7417 &  
$\frac{1}{2}$ & \,$\frac{1}{2}$ &10 \\[0.07cm]
$\Lambda$(1405) & 10.42 &  0,\,$\frac{1}{2}$ & 10 & & $\Lambda^+_c$ & 
16.928 & 
0 & \,$\frac{1}{2}$ & 17 \\[0.07cm]
$\Lambda$(1670) & 12.37 & 0,\,$\frac{1}{2}$ & 12 & &  $\Lambda_c$(2593) & 
19.215 &  0 & \,$\frac{1}{2}$ & 19  \\[0.07cm] 
$\Lambda$(1800) & 13.33 & 0,\,$\frac{1}{2}$ & 13 & & $\Sigma^0_c$  & 
18.167 & 
1 & \,$\frac{1}{2}$ & 18 \\[0.07cm]
$\Sigma^0$ & 8.8352 & 1,\,$\frac{1}{2}$ & 9 & & $\Xi^0_c$ & 18.302 & 
$\frac{1}{2}$ & \,$\frac{1}{2}$ &18 \\[0.07cm]
$\Sigma$(1660)  & 12.298 & 1,\,$\frac{1}{2}$  & 12  & &  $\Omega^0_c$  & 
20.03 & 0 & \,$\frac{1}{2}$ & 20 \\[0.07cm]
$\Sigma$(1750)  & 12.965  & 1,\,$\frac{1}{2}$ & 13 \\
[.2cm]\hline\hline
	\end{tabular}
	\end{table}

	\begin{table}[h]\caption{The $\gamma$-branch mesons with I,J=0,0}
 	\begin{tabular}{lllllrcc||cclll}\\
\hline\hline\\
& & & particle  &  m/m($\pi^0$) & N  & &   &  & particle & m/m($\pi^0$) & 
N\\ 
[0.5ex]\hline
\\[-0.3cm]
& & & $\eta$  &  4.0559  &  4 &   & &  &  $\eta(1440)$  & 10.48  & 10  
\\[0.07cm]
& & & $\eta'$  &  7.0958  & 7 &  & & &  $f_0$(1500)  &  11.135  & 
11 \\[0.07cm]
& & & $f_0$  &  7.261 &  7 & & & &  $\eta_c$  &  22.076 &  22 \\[0.07cm]
& & & $\eta$(1295)  &  9.594 &  10 & & & &  $\chi_{c_0}$  &  25.301 &  
25\\ 
[0.2cm]\hline\hline
	\end{tabular}
	\end{table}
\noindent 
The $\Omega^-$ particle has been omitted from Table 2 because it 
has spin 3/2, and the $\Lambda$(1810) resonance has also been omitted 
because it differs from $\Lambda$(1800) only in parity, as have been two 
$\Xi$  resonances whose spin is uncertain. A least square analysis of the 
masses in Table 2 yields the formula

\begin{equation}
\mathrm{m}(N)/\mathrm{m}(\pi^0) = 1.0013\,N + 0.1259 \qquad    N > 1,   
\end{equation}

\noindent
with the very good correlation coefficient 0.997, and the masses in Table 
3 are described by the equation

\begin{equation} \mathrm{m}(N)/\mathrm{m}(\pi^0) = 1.0055\,N + 0.0592 \qquad 
N >1 ,
\end{equation}
\noindent
with r$^2$ = 0.999. If we combine the 
particles in Tables 1,2,3, that means if we consider all mesons and 
baryons of the $\gamma$-branch, ``stable" or unstable, with I $\leq$ 1, J 
$\leq$ 1/2 then we obtain from a least square analysis the formula 

\begin{equation}    \mathrm{m}(N)/\mathrm{m}(\pi^0) = 1.0056\,N + 0.0610\qquad    
N \ge 1,  
\end{equation}

\noindent
with the correlation coefficient 0.9986.
 
	\vspace{0.5cm}
	\includegraphics{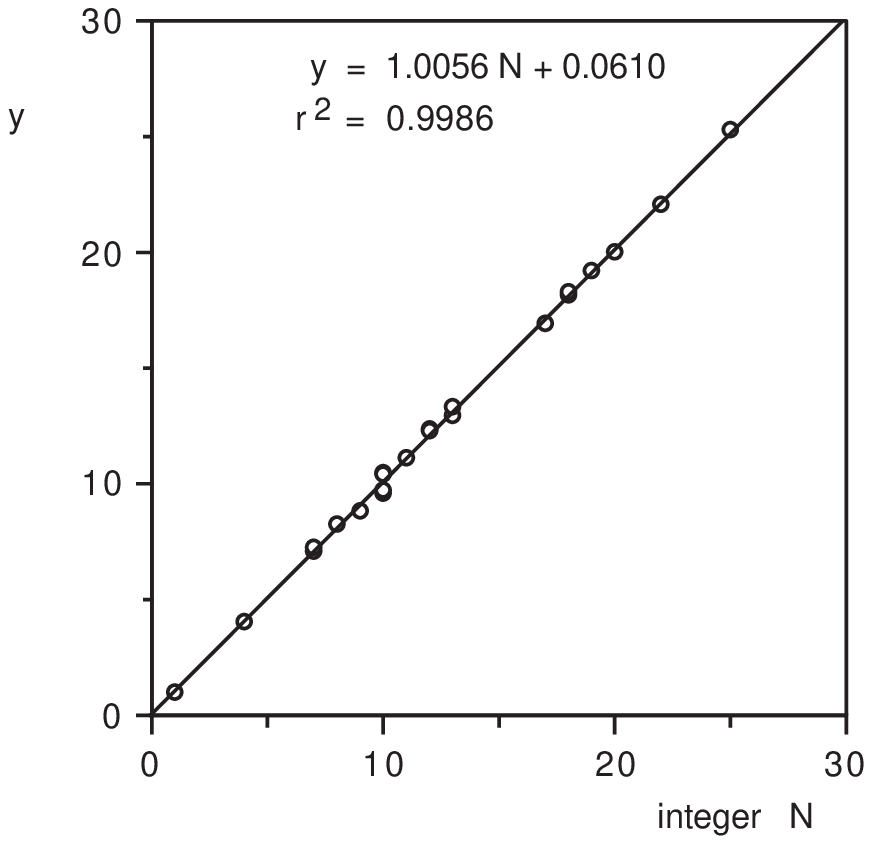}
	\vspace{-0.3cm}
	\begin{quote}
Fig.\,1: The mass of all mesons and baryons with 
I\,$\leq$ 1,\,J\,$\leq\frac{1}{2}$ in units of m($\pi^0$) as a function of 
the integer  N. y = m/m($\pi^0$)
	\end{quote}

\noindent
The line through the points is 
shown in Fig.\,1 which tells that 22 particles of the $\gamma$-branch of 
different spin and isospin, strangeness and charm; eight I,J = 0,0 mesons, thirteen \,\,J = 1/2 
baryons and the $\pi^0$ meson with I,J = 1,0, lie on a straight line with 
slope 1.0056. In other words they approximate the integer multiple rule 
very well.

   Searching for what else the 
$\pi^0$,\,$\eta$,\,$\Lambda$,\,$\Sigma^0$,\,$\Xi^0$,\,$\Omega^-$ particles 
have 
in common, we find that the principal decays (decays with a fraction $>\,1\%$) 
of these particles, as listed in Table 1, involve primarily $\gamma$-rays, 
the characteristic case is $\pi^0 \rightarrow \gamma\gamma$  (98.8\%). We 
will later on discuss a possible explanation for the 1.198\% of the decays 
of $\pi^0$ which do not follow the $\gamma\gamma$ route. After the 
$\gamma$-rays the next most 
frequent decay product of the heavier particles of the $\gamma$-branch are 
$\pi^0$ mesons which again decay into $\gamma\gamma$. To describe the 
decays in another way, the principal decays of the particles listed above 
take place $\emph{always without the emission of neutrinos}$\,; see Table 
1. 
There the decays and the fractions of the principal decay modes are given, 
taken from the Particle Physics Summary. We cannot consider decays 
with fractions $< 1\%$. We will refer to the particles whose masses are 
approximately integer multiples of the mass of the $\pi^0$ meson, and 
which decay without the emission of neutrinos, as the 
$\gamma$-$\emph{branch}$ of the particle spectrum.

   To summarize the facts concerning the $\gamma$-branch. Within 
0.66\% on the average 
the masses of the particles of the $\gamma$-branch are integer multiples 
(namely 4,\,8,\,9,\,10,\,12, and even 17,\,18,\,20) of the mass of the 
$\pi^0$ meson. 
It is improbable that nine particles have masses so close to integer 
multiples of m($\pi^0$) if there is no correlation between them and the 
$\pi^0$ meson. It has, on the other hand, been argued that the integer 
multiple rule is a numerical coincidence. But the probability that the 
mass ratios of the $\gamma$-branch fall by coincidence on integer numbers 
between 1 and 20 instead on all possible numbers between 1 and 20 with two 
decimals after the period is smaller than 10$^{-20}$, i.e nonexistent. The 
integer multiple rule is not affected by more than 3\% by the spin, the 
isospin, the strangeness, and by charm. The integer multiple rule seems 
even to apply to the $\Omega^-$ and $\Lambda_c^+$ particles, although they 
are charged. In order for the integer multiple rule to be valid the 
deviation of the ratio m/m($\pi^0$) from an integer number must be smaller 
than 1/2N, where N is the integer number closest to the actual ratio 
m/m($\pi^0$). That means that the permissible deviation decreases rapidly 
with increased N. All particles of the $\gamma$-branch have deviations smaller than 
1/2N.
 
   The remainder of the stable mesons and baryons are the  
$\pi^\pm$,\, K$^{\pm,0}$,\,p,\,\,n,\quad D$^{\pm,0}$, and D$_S^\pm$ 
particles which make up 
the \emph{$\nu$-branch} of the particle spectrum. The ratios of the masses are given in Table 
4.

	\begin{table}\caption{The $\nu$-branch of the particle 
spectrum}
	\begin{tabular}{llllrcl}\\

\hline\hline\\
 & m/m($\pi^\pm$) & multiples & decays\footnotemark[1]
 & fraction & spin & mode 
 \\
 & & & & (\%) & & \\
[0.5ex]\hline
\\
$\pi^\pm$ & 1.0000 & 1.0000\,\,$\cdot$\,\,$\pi^\pm$ & $\mu^+\nu_\mu$ & \,\,99.9877 & 0 & (1.1)\\
\\
K$^{\pm,0}$ & 3.53713 & 0.8843\,$\cdot$\,\,4$\pi^\pm$ & $\mu^+\nu_\mu$ & 
\,\,63.57 & 0 & 
(2.2) + $\pi^0$\\
 & & & $\pi^+\pi^0$ & \,\,21.16 & &\\
 & & & $\pi^+\pi^-\pi^+$ & \,\,\,\,\,5.59 & &\\
 & & & $\pi^0 e^+ \nu_e$ & \,\,\,\,\,4.82 & &\\
 & & & $\pi^0\mu^+\nu_\mu$ & \,\,\,\,\,3.18 & &\\
\\
n & 6.73186 & 0.8415\,$\cdot$\,\,8$\pi^\pm$ & p\,e$^-\overline{\nu}_e$ & 
100.\,\,\,\,\,\,\,\,  & $\frac{1}{2}$ & 2$\cdot$((2.2) + $\pi^0$)\\
 & & 0.9617\,$\cdot$\,\,7$\pi^\pm$\\
 & & 0.9516\,$\cdot$\,\,2$K^\pm$ \\
\\
D$^{\pm,0}$ & 13.393 & 0.8370\,$\cdot$\,16$\pi^\pm$ & $e^+$ anything & \,\,17.2 & 
0 & 2(2$\cdot((2.2) + \pi^0)$)\\
 & & 0.9466\,$\cdot\,\,4K^\pm$ & K anything & \,\,24.2 \\
 & & 0.9954\,$\cdot$\,(n+p) & $\overline{\mathrm{K}}^0$ anything\\	
 & &                       & \,\,+K$^0$ anything & \,\,59\\
 & &                       & $\eta$ anything & $<$\,13\\
\\
D$^\pm_S$ & 14.104 & 0.8296\,$\cdot$\,17$\pi^\pm$ & K anything & \,\,13 & 0 & 
2(2$\cdot$((2.2) + $\pi^0$))\\
 & & 0.9968\,$\cdot$\,\,4K$^\pm$ & $\overline{\mathrm{K}}^0$ 
 anything & & & \,\, + $\pi^0$\\
 & &                            & \,\,+K$^0$ anything & \,\,39\\
 & &                            & K$^+$ anything & \,\,20\\
 & &                            & e$^+$ anything & $<20$\\
[0.2cm]\hline\hline

\vspace{0.1cm}
	\end{tabular}

	\footnotemark{\footnotesize The particles with negative charges have conjugate charges of the listed  decays.}

	\end{table}

These particles are in general charged, exempting the K$^0$ and D$^0$ 
mesons and the neutron n, in contrast to the particles of the 
$\gamma$-branch, which are in general neutral. It does not make a 
significant difference whether one considers the mass of a particular 
charged or neutral particle. After the $\pi$ mesons, the largest mass 
difference between charged and neutral particles is that of the K mesons 
(0.81\%), and thereafter all mass differences between charged and neutral 
particles are $< 0.5\%$. The integer multiple rule does not immediately 
apply to the masses of the $\nu$-branch particles if m($\pi^\pm$) (or 
m($\pi^0$)) is used as reference, because m(K$^\pm$) = 
0.8843\,$\cdot$\,4m($\pi^\pm$). 0.8843\,$\cdot$\,4 = 3.537 is far from 
integer. Since 
the masses of the $\pi^0$ meson and the $\pi^\pm$ mesons differ by only 
3.4\% it has been argued that the $\pi^\pm$ mesons are, but for the 
isospin, the same particles as the $\pi^0$ meson, and that therefore the 
$\pi^\pm$ cannot start another particle branch. However, this argument is 
not supported by the completely different decays of the $\pi^0$ mesons and 
the $\pi^\pm$ mesons. The $\pi^0$ meson decays almost exclusively into 
$\gamma\gamma$ (98.8\%),  whereas the $\pi^\pm$ mesons decay practically 
exclusively into $\mu$-mesons and neutrinos, e.g. $\pi^+$ $\rightarrow$  
$\mu^+$  + $\nu_\mu$  (99.9877\%). Furthermore, the lifetimes of the 
$\pi^0$  and the $\pi^\pm$ mesons differ by nine orders of magnitude, 
being $\tau$($\pi^0$) = 8.4\,$\cdot$\,10$^{-17}$ sec versus 
$\tau$($\pi^\pm$) 
= 2.6\,$\cdot$\,10$^{-8}$ sec.

   If we make the $\pi^\pm$ mesons the reference particles of the 
$\nu$-branch, then we must multiply the mass ratios m/m($\pi^\pm$) of the 
above listed particles with an average factor 0.848 $\pm$ 0.025, as follows from 
the mass ratios on Table 4. 

\noindent
The integer multiple rule may, however, apply 
directly if one makes m(K$^\pm$) the reference for masses larger than 
m(K$^\pm$). The mass of the neutron is 0.9516\,$\cdot$\,2m(K$^\pm$), which 
is 
only a fair approximation to an integer multiple. There are, on the other 
hand, outright integer multiples in m(D$^\pm$) = 0.9954\,$\cdot$\,(m(p) + 
m(n)), and in m(D$_S^\pm$) = 0.9968\,$\cdot$\,4m(K$^\pm$). A least square 
analysis of the masses of the $\nu$-branch in Table 4 yields the formula

\begin{equation} \mathrm{m}(N)/0.853\mathrm{m}(\pi^\pm) = 1.000\,N + 
0.00575\qquad  N \ge 1 ,
\end{equation}
\noindent
with r$^2$ = 0.998. This means that the particles of the $\nu$-branch are 
integer multiples of m($\pi^\pm$) times the factor 0.853. One must, 
however, consider that the $\pi^\pm$ mesons are not necessarily the 
perfect reference for all $\nu$-branch particles, because $\pi^\pm$ has I 
= 1, whereas for example K$^\pm$ has I = 1/2 and S = $\pm$1 and the 
neutron has also I = 1/2. Actually the 
factor 0.853 is only an average. The mass ratios indicate that the factor 
before m/m($\pi^\pm$) decreases slowly with increased N. The existence of the factor 
and its decrease will be explained later.

   Contrary to the particles of the $\gamma$-branch, the $\nu$-branch 
particles decay preferentially with the emission of neutrinos, the 
foremost example is, e.g., $\pi^+ \rightarrow \mu^+$ + $\nu_\mu$ with a 
fraction of 99.988\%. Neutrinos characterize the weak interaction. We will 
refer to the particles in Table 4 as the $\emph{neutrino branch}$ 
($\nu$-branch) of the particle spectrum. We emphasize that a weak decay of 
the particles of the $\nu$-branch is by no means guaranteed. Although the 
neutron decays via n $\rightarrow$ p + e$^-$ + $\bar{\nu}_e$ in 887 sec 
(100\%), the proton is stable. There are, on the other hand, decays as e.g. K$^+ 
\rightarrow \pi^+\pi^-\pi^+$ (5.59\%), but the subsequent decays of the 
$\pi^\pm$ mesons lead to neutrinos and e$^\pm$. The decays of the particles in the 
$\nu$-branch follow a mixed rule, either weak or electromagnetic.

   To summarize the facts concerning the $\nu$-branch of the mesons and 
bary-ons. The masses of these particles seem to follow the integer 
multiple 
rule if one uses the $\pi^\pm$ meson as reference, however the mass ratios 
share a common factor 0.85 $\pm$ 0.025.

   To summarize what we have learned about the integer multiple rule: In 
spite of differences in charge, spin, strangeness, and 
charm the masses of the mesons and baryons of the $\gamma$-branch are 
 integer multiples 
of the mass of the $\pi^0$ meson within at most 3.3\% and on the average within 0.66\%. Correspondingly, the masses of the 
particles of the $\nu$-branch are, after multiplication with a factor 0.85 
$\pm$ 0.025, integer multiples of the mass of the $\pi^\pm$ mesons. The 
validity of the integer multiple rule can easily be verified with a 
calculator from the data 
in the 
Particle Physics Summary. The integer multiple rule 
suggests that the particles are the result of superpositions of modes and 
higher modes of a wave equation.

\section {Standing waves in a cubic lattice and the particles of the 
$\gamma$-branch} 

  In an earlier article [5] we have tried to explain the masses of the 
elementary particles by monochromatic eigenfrequencies of standing waves 
in an elastic cube. The explanation of the particles by monochromatic 
eigenfrequencies does not seem to be tenable because a monochromatic 
frequency does not accommodate the multitude of frequencies created in a 
high energy collision of 10$^{-23}$ sec duration. We will now study, as we 
have done in [6], whether the so-called stable particles of the 
$\gamma$-branch cannot be described by the frequency spectrum of standing 
waves in a cubic lattice, which can accommodate automatically the Fourier 
frequency spectrum of an extreme short-time collision. 
   The investigation of the consequences of lattices for particle theory 
was initiated by Wilson [7] who studied a fermion lattice. This study has 
developed over time into lattice QCD. The results of such endeavors have 
culminated in the paper of Weingarten [8] and his colleagues which 
required elaborate year long numerical calculations. They determined the 
masses of seven particles 
(K$^*$,p,$\phi$,$\Delta$,$\Sigma$,$\Xi$,$\Omega$), with an uncertainty of 
up to $\pm$8\%, agreeing with the observed masses within a few percent, 
up to 6\%. Our theory covers all particles of the $\gamma$-branch, namely 
the 
$\pi^0$,\,$\eta$,\,$\Lambda$,\,$\Sigma^0$,\,$\Xi^0$,\,$\Omega^-$,\,$\Lambda_c^+$,\,$\Sigma_c^0$,\,\,$\Xi_c^0$ 
and $\Omega_c^0$ particles and agrees on the average with the measured 
masses within 0.66\%, and covers also the particles of the $\nu$-branch, 
as we will see later.

   It will be necessary to outline the most elementary aspects of the 
theory of lattice oscillations since we will, in the following, 
investigate whether the masses of the $\gamma$-branch particles can be 
explained with the help of the frequency spectra of standing waves in a 
cubic lattice. The classic paper describing lattice oscillations is from 
Born and v.Karman [9], henceforth referred to as B\&K. They looked at 
first 
at the oscillations of a one-dimensional chain of points with mass m, 
separated by a constant distance $\emph{a}$. This is the 
$\emph{monatomic}$ case, all lattice points have the same mass. B\&K 
assume 
that the forces exerted on each point of the chain originate only from the 
two neighboring points. These forces are opposed to and proportional to 
the displacements, as with elastic springs. The equation of motion is in 
this case

\begin{equation} m\ddot{u}_{n} = \alpha(u_{n+1} - u_n) - \alpha(u_n - 
u_{n-1})\,.
\end{equation}
\noindent
The $u_n$ are the displacements of the mass points from their equilibrium position 
which are apart by the distance \emph{a}. The dots signify, as usual, differentiation with respect 
to time, $\alpha$ is a constant characterizing the force between the 
lattice points, and n is an integer number.
 
   In order to solve (6) B\&K set 

\begin{equation} u_n = Ae^{i(\omega\,t\, +\, n\phi)}\,,
\end{equation}

\noindent
which is obviously a temporally and spatially periodic solution. n is an 
integer, with n\,$\leq$\,N, where N is the number of points in the chain. 
$\phi$ = 0 is the monochromatic case.  At n$\phi$ = $\pi$/2 there are 
nodes, where for all times $\emph{t}$ the displacements are zero, as with standing 
waves. If a displacement is repeated after n points we have n$\emph{a}$ = 
$\lambda$, where $\lambda$ is the wavelength, $\emph{a}$ the lattice 
constant, and it must be n$\phi$ = 
2$\pi$ according to (7). It follows that

\begin{equation} \lambda = 2\pi\emph{a}/\phi\,.
\end{equation}
 
Inserting (7) into (6) one obtains a continuous frequency spectrum given by 
Eq.(5) of B\&K

\begin{equation}  \omega = 2\sqrt{\alpha/m}\,sin(\phi/2)\,.
\end{equation}
\noindent
B\&K point out that there is not only a continuum of frequencies, but also 
a maximal frequency which is reached when $\phi$ = $\pi$, or at the 
minimum of the possible wavelengths $\lambda$ = 2$\emph{a}$.
   B\&K then discuss the three-dimensional lattice, with lattice constant 
$\emph{a}$ and masses m. They reduce the complexities of the problem by 
considering only the 18 points nearest to any point. These are 6 points at 
distance $\emph{a}$, and 12 points at distance $\emph{a}\sqrt{2}$. B\&K 
assume that the forces between the points are linear functions of the 
small displacements, that the symmetry of the lattice is maintained, and 
that the equations of motion transform into the equations of motion of 
continuum mechanics for $\emph{a}$ $\rightarrow$  0. We cannot reproduce 
the lengthy equations of motion of the three-dimensional lattice. In the 
three-dimensional case we deal with the forces caused by the 6 points at 
the distance $\emph{a}$ which are characterized by the constant $\alpha$ 
in the case of central forces. There are also the forces which originate 
from the 12 points at distance $\emph{a}\sqrt{2}$, characterized by the 
constant $\gamma$, which we do not need. We investigate the 
propagation of plane waves in a three-dimensional monatomic lattice with 
the ansatz

\begin{equation} u_{l,m} = u_0\,e^{i(\omega\,t\, + \,l\phi_1\, +\, 
m\phi_2)}\,,
\end{equation}

\noindent
and a similar ansatz for $v_{l,m}$, with l,m being integer numbers $\leq 
N^\frac{1}{3}$, where $N^\frac{1}{3}$ is the number of lattice points 
along a side of the cube. We also consider higher order solutions, with 
$i_1\cdot$l and $i_2\cdot$m, where $i_1,i_2$ are integer numbers. The 
boundary conditions are periodic. The number of normal modes must be equal 
to the number of particles in the lattice. B\&K arrive, in the case of 
two-dimensional waves, at a secular equation for the frequencies

\begin{equation}
\left| \begin{array}{cc}
A(\phi_1,\phi_2) - m\nu^2 & B(\phi_2,\phi_1)\\
B(\phi_1,\phi_2) & A(\phi_2,\phi_1) - m\nu^2
\end{array}\right| = 0\,.
\end{equation}
\noindent
The formulas for A($\phi_1,\phi_2$) and B($\phi_1,\phi_2$) are given in 
equation (17) of B\&K.
 
    The theory of lattice oscillations has been pursued in particular by 
Blackman [10], a summary of his and other studies is in [11]. 
Comprehensive reviews of the results of linear studies of lattice dynamics 
have been written by Born and Huang [12], by Maradudin et al. [13], and by 
Ghatak and Kothari [14].

\section {The masses of the particles of the $\gamma$-branch}

   We will now assume, as seems to be quite natural, that the particles of 
the $\gamma$-branch $\emph{consist of the same particles into which they 
decay}$, directly or ultimately. That means that the $\gamma$-branch 
particles consist of photons. We base this assumption on the fact that photons 
and $\pi^0$ mesons are the principal mode of decay of the $\gamma$-branch particles, the 
characteristic example is $\pi^0$ $\rightarrow$ $\gamma\gamma$  (98.8\%). 
Table 1 shows that there are decays of the $\gamma$-branch particles which 
lead to particles of the $\nu$-branch, in particular to pairs of $\pi^+$ 
and $\pi^-$ mesons, already present in 28\% of the decays of the $\eta$ 
meson, as in $\eta \rightarrow \pi^+\pi^-\pi^0$ (23.2\%). It appears 
that this has to do with pair production in the $\gamma$-branch particles. 
Pair production is evident in the decay $\pi^0 \rightarrow e^+ + e^- + 
\gamma$ (1.198\%). It requires the presence of electromagnetic waves of high 
energy. Anyway, the explanation of the $\gamma$-branch particles must begin 
with the explanation of the most simple example of its kind, the $\pi^0$ 
meson, which by all means seems to consist of photons. 
The composition of the particles of the $\gamma$-branch suggested here 
offers a direct route from the formation of a $\gamma$-branch particle, 
through its lifetime, to its decay products. Particles that are made of 
photons are necessarily neutral, as the majority of the particles of the 
$\gamma$-branch are.
 
   We also base our assumption that the particles of the $\gamma$-branch 
are made of photons on the circumstances of the formation of the 
$\gamma$-branch particles. The most simple and straightforward creation of 
a $\gamma$-branch particle are the reactions $\gamma$ + p $\rightarrow$ 
$\pi^0$ + p, or in the case that the spins of $\gamma$ and p are parallel $\gamma$ + p $\rightarrow$ 
$\pi^0$ + p + $\gamma^\prime$. A photon impinges on a proton and creates a 
$\pi^0$ meson. In a timespan on the order of 10$^{-23}$ sec the pulse of the 
incoming electromagnetic wave is, according to Fourier analysis, converted 
into a continuum of electromagnetic waves with frequencies ranging from 
10$^{23}$ sec$^{-1}$ to $\nu$ $\rightarrow$ $\infty$. The wave packet so 
created decays, according to experience, after 8.4\,$\cdot$\,10$^{-17}$ sec 
into two electromagnetic waves or $\gamma$-rays. It seems to be very 
unlikely that Fourier analysis does not hold for the case of an 
electromagnetic wave impinging on a proton. The question then arises of 
what happens to the electromagnetic waves in the timespan of 10$^{-16}$ 
seconds between the creation of the wave packet and its decay into two 
$\gamma$-rays? We will investigate whether the electromagnetic waves 
cannot continue to exist for the 10$^{-16}$ seconds until the wave packet 
decays.

   There must be a mechanism which holds the wave packet of the newly 
created particle together, or else it will disperse. We assume that the 
very many photons in the new particle are held together in a cubic 
lattice. Ordinary cubic lattices, such as the NaCl lattice, are held 
together by attractive forces between particles of opposite polarity. We 
assume that the photon lattice is held together by weak attractive forces 
between photons, if they are about 10$^{-16}$ cm apart. Electrodynamics 
does not predict the existence of such a force between two photons. 
However, electroweak theory says that $e^2$ $\approx$ $g^2$ on the scale 
of 10$^{-16}$ cm, and we will now assume that there is such a weak force 
in the photon lattice. The potential of this force is given in section 8, 
Eq.(45). It is not unprecedented that photons have been 
considered to be building blocks of the elementary particles. Schwinger 
[15] has once studied an exact one-dimensional quantum electrodynamical 
model in which the photon acquired a mass $\sim$ $e^2$.

   We will now investigate the standing waves of a cubic photon lattice. 
We assume that the lattice is held together by a weak force acting from 
one lattice point to the next. We assume that the range of this force is 
10$^{-16}$ cm, because the range of the weak nuclear force is on the order 
of 
10$^{-16}$ cm, according to [16]. For the sake of simplicity we set the 
sidelength of the lattice at 10$^{-13}$ cm, the exact size of the nucleon 
is given in [17] and will be used later. With \emph{a} = $10^{-16}$ cm there are then 10$^9$ lattice 
points. As we will see the ratios of the masses of the particles are independent of the 
sidelength of the lattice. Because it is the most simple case, we assume 
that a central force acts between the lattice points. We cannot consider 
spin, isospin, strangeness or charm of the particles. The frequency 
equation for the oscillations of an isotropic monatomic cubic lattice with 
central forces is either given by equation (9), or in the two-dimensional 
case by

\begin{equation} 4\pi^2\nu^2 = \frac{2\alpha}{M}(2 - cos\phi_1cos\phi_2 + 
sin\phi_1sin\phi_2 - cos\phi_1)\,.
\end{equation} 

   According to Eq.(13) of B\&K

\begin{equation}   \alpha = \emph{a}(c_{11} - c_{12} - c_{44})\,,
\end{equation}

\noindent
where c$_{11}$, c$_{12}$ and c$_{44}$ are the elastic constants in 
continuum mechanics which applies in the limit $\emph{a}$ $\rightarrow$  
0. If we consider central forces then c$_{12}$ = c$_{44}$ which is the 
classical Cauchy relation. Isotropy requires that c$_{44}$ = (c$_{11}\,-\,$ 
c$_{12}$)/2. Eq.(12) follows directly from 
the equation of motion for the displacements in a monatomic lattice, which 
are given e.g. by Blackman [10]. The minus sign in front of cos$\phi_1$ in 
(12) 
means that the waves are longitudinal. Transverse waves in a cubic lattice 
with concentric forces are not possible according to [14]. All frequencies 
that solve (12) come with either a plus or a minus sign which is, as we 
will see, important.

   The frequency distribution following from (12) is shown in Fig.\,2. 

	\vspace{0.5cm}
	\hspace{2.5cm}
	\includegraphics{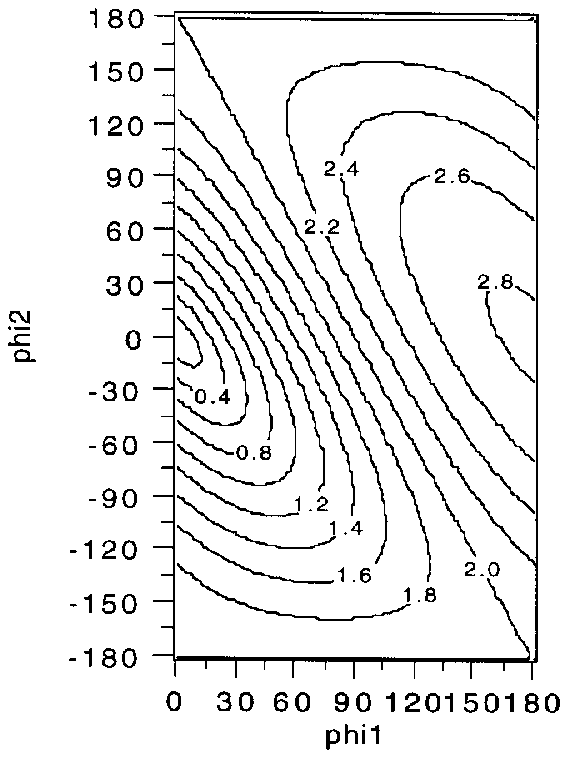}
	\vspace{-0.3cm}
	\begin{quote}
Fig.\,2: The frequency distribution $\nu/\nu_0$  of the basic mode according 
to Eq.(12) in units of $\nu_0$. Monatomic, isotropic case.
	\end{quote}

There are some easily verifiable frequencies. For example at 
$\phi_1,\phi_2$ = 0,0 it is $\nu$ = 0, at $\phi_1,\phi_2$ = $\pi/2,\pi/2$ 
it is $\nu = \nu_0\sqrt{6}$, at $\phi_1,\phi_2$ = $\pi/2,-\pi/2$ we have 
$\nu = \nu_0\sqrt{2}$. Furthermore at $\phi_1,\phi_2$ = $\pi$,0 we have 
$\nu = \nu_0\sqrt{8}$, and for all values of $\phi_1$ it is $\nu = 2\nu_0$ 
at $\phi_2$ = $\pi$ and $\phi_2$ = $-\pi$, with

\begin{equation}  \nu_0 = \sqrt{\alpha/4\pi^2M}\,,
\end{equation}

\noindent
or as we will see, using Eq.(16),  $\nu_0$  =  c$_\star/2\pi\emph{a}$.

   The limitation of the group velocity in the photon lattice has now to 
be considered. The group velocity is given by

\begin{equation}  c_g = \frac{d\omega}{dk} = 
\emph{a}\sqrt{\frac{\alpha}{M}}\cdot\frac{df(\phi_1,\phi_2)}{d\phi}\,.
\end{equation}
\noindent
The group velocity in the photon lattice has to be equal to the velocity 
of light c$_\star$ throughout the entire frequency spectrum, because 
photons move with the velocity of light. In order to learn how this 
requirement affects the frequency distribution we have to know the value 
of $\sqrt{\alpha/M}$ in a photon lattice. But we do not have information 
about what either $\alpha$ or M might be in this case. We assume in the 
following that $\emph{a}\sqrt{\alpha/M}$ = c$_\star$, which means, since 
$\emph{a} = 10^{-16}$ cm, that $\sqrt{\alpha/M}$ = 3\,$\cdot$\,10$^{26}$ 
sec$^{-1}$, or that the corresponding period is $\tau$ = 
1/3\,$\cdot$\,10$^{-26}$ sec, which is the time it takes for a wave to 
travel 
with the velocity of light over one lattice distance. With 

\begin{equation} c_\star = \emph{a}\sqrt{\alpha/M}\,
\end{equation}

\noindent
the equation for the group velocity is

\begin{equation} c_g = c_\star\cdot df/d\phi\,.
\end{equation} 

  For a photon lattice that means, since c$_g$ must then always be equal 
to c$_\star$, that df/d$\phi$ = 1. This requirement determines the form of 
the frequency distribution, whether we deal with an axial oscillation as 
in Eq.(9) or the two-dimensional oscillation of Eq.(12), regardless of the 
order of the mode of oscillation. The frequencies of the corrected 
spectrum must increase from $\nu$ = 0 at the origin $\phi_1,\phi_2$ = 0,0 with slope 1 
(in units of $\nu_0$) until a maximum is reached, from where the frequency 
must decrease with slope 1 to $\nu$ = 0. The frequency distribution 
corrected for (17), i.e. with $df/d\phi$ = 1, is shown in Fig.\,3. The corrected frequency 
distributions of higher modes are of the same type, but for the area they 
cover, see e.g. Fig.\,4.
 
	\begin{figure}[h]
	\unitlength1cm
	\begin{minipage}[t]{6.5cm}
	\includegraphics{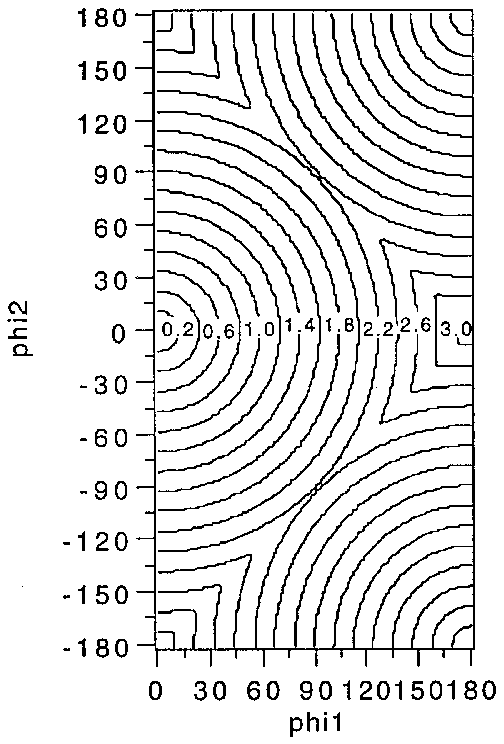}
	\begin{quote}
Fig.\,3: The frequency distribution $\nu/\nu_0$ of the basic (1.1) mode 
with df/d$\phi$ = 1.
	\end{quote}
	\end{minipage}
	  \hfill
	\begin{minipage}[t]{6.5cm}
	\includegraphics{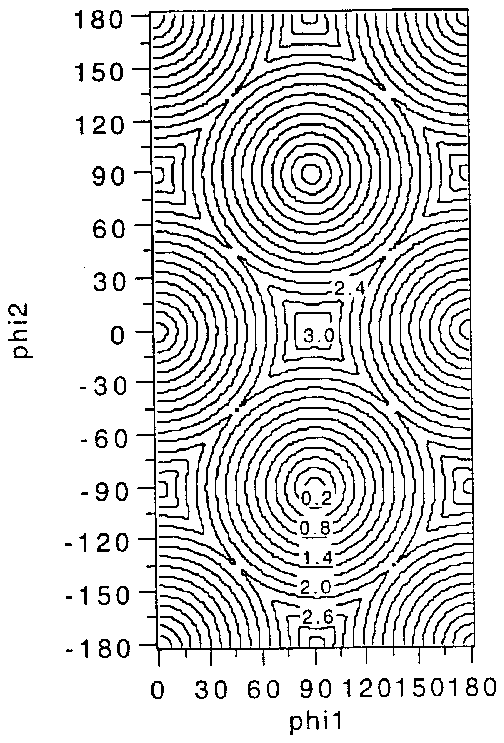}
	\begin{quote}
Fig.\,4: The frequency distribution $\nu/\nu_0$ of the (2.2) mode, 
with df/d$\phi$ = 1. The variables $\phi_1$ and 
$\phi_2$ are one-half of the actual $\phi$ of the second mode.
	\end{quote}
	\end{minipage}
	\end{figure}

The second mode ($i_1,i_2$ = 2) covers 4 times 
the 
area of the basic mode, because 2$\phi$ ranges from 0 to 2$\pi$, (for 
$\phi$ $>$ 0), whereas the basic mode ranges from 0 to $\pi$. Consequently 
the energy (Eq.18) contained in all frequencies of the second mode is 
four times larger than the energy of the basic mode, because the energy 
contained in the lattice oscillations must be proportional to the sum of 
all frequencies. Adding, by superposition, to the second mode different 
numbers of basic modes or of second modes will give exact integer multiples 
of the energy of the basic mode. Now we understand the integer multiple 
rule of the particles of the $\gamma$-branch. There is, in the framework 
of this theory, on account of Eq.(17), no alternative but $\emph{integer 
multiples}$ of the basic mode for the energy contained in the frequencies 
of the different modes or for superpositions of different modes. In other 
words, the masses of the different particles are integer multiples of the 
mass of the $\pi^0$ meson, assuming that there is no spin, isospin, 
strangeness or charm.

   We remember that the measured masses in Table 1, 
which incorporate different spins, isospins, strangeness, and charm spell 
out the integer multiple rule within on the average 0.66\% accuracy. It is worth noting 
that $\emph{there is no free parameter}$ if one takes the ratio of the 
energies contained in the frequency distributions of the different modes, 
because the factor $\sqrt{\alpha/M}$ in Eqs.(9,12) cancels. This means, in 
particular, that the ratios of the frequency distributions, or the mass 
ratios, are independent of the mass of the photons at the lattice points, 
as well as of the magnitude of the force between the lattice points.

   It is obvious that the integer multiples of the frequencies are only a 
first approximation of the theory of lattice oscillations and of the mass 
ratios of the particles. The equation of motion in the lattice (6) does 
not apply in the eight corners of the cube, nor does it 
apply to the twelve edges, nor, in particular, to the six sides of the 
cube. A cube with 10$^9$ lattice points is not correctly described by the 
periodic boundary condition we have used, but is what is referred to as a 
microcrystal. A phenomenological theory of the frequency distributions in 
microcrystals, considering in particular surface waves, can be found in 
Chapter 6 of Ghatak and Kothari [14]. Surface waves may account for the 
small deviations of the mass ratios of the mesons and baryons from the 
integer multiple rule of the oscillations in a cube. However, it seems to be futile to 
pursue a more accurate determination of the oscillation frequencies as 
long as one does not know what the structure of the electron is, whose 
mass is 0.378\% of the mass of the $\pi^0$  meson and hence is a 
substantial part of the deviation of the mass ratios from the integer 
multiple rule.
 
   Let us summarize our findings concerning the particles of the 
$\gamma$-branch. The $\pi^0$ meson is the basic mode of the photon lattice 
oscillations. The $\eta$ meson corresponds to the first higher mode 
($i_1,i_2$ = 2), as is suggested by m($\eta$) $\approx$ 4m($\pi^0$). The 
$\Lambda$ particle corresponds to the superposition of two higher modes 
($i_1,i_2 = 2$), as is suggested by m($\Lambda) \approx$ 2m($\eta$). This 
superposition apparently results in the creation of spin 1/2. The two 
modes would then have to be coupled. The $\Sigma^0$ and $\Xi^0$ baryons 
are superpositions of one or two basic modes on the $\Lambda$  particle. 
The $\Omega^-$ particle corresponds to the superposition of three coupled 
higher modes ($i_1,i_2$ = 2) as is suggested by m($\Omega^-$) $\approx$ 
3m($\eta$). This procedure apparently causes spin 3/2. The charmed 
$\Lambda_c^+$ particle seems to be the first particle incorporating a 
(3.3) mode. $\Sigma_c^0$ is apparently the superposition of a basic mode 
on $\Lambda_c^+$, as is suggested by the decay of $\Sigma_c^0$. The 
easiest explanation of $\Xi_c^0$ is that it is the superposition of two 
coupled (3.3) modes. The superposition of two modes of the same type is, 
as in the case of $\Lambda$, accompanied by spin 1/2. The $\Omega_c^0$ 
baryon is apparently the superposition of two basic modes on the $\Xi_c^0$ 
particle. All neutral particles of the $\gamma$-branch are thus accounted 
for. We find it interesting that all $\gamma$-branch particles with 
coupled 2$\cdot$(2.2) modes, or the $\Omega^-$ particle with the coupled 
3$\cdot$(2.2) 
mode, have strangeness. But this rule does not hold in the presence of a 
(3.3) mode. All $\gamma$-branch particles with a (3.3) mode have charm. 
The modes of the particles are listed in Table 1.

   We have also found the $\gamma$-branch $\emph{antiparticles}$, which 
follow from the negative frequencies which solve Eqs.(9) or (12). 
Antiparticles have always been associated with negative energies. 
Following Dirac's argument for electrons and positrons, we associate the 
masses with the negative frequency distributions with 
antiparticles. We emphasize that the existence of antiparticles is an 
automatic consequence of our theory.

   All particles of the $\gamma$-branch are unstable with lifetimes on the 
order of 10$^{-10}$ sec or shorter. Born [18] has shown that the 
oscillations in cubic lattices held together by central forces are 
unstable. It also seems to be possible to understand the decay of the 
$\pi^0$  meson $\pi^0 \rightarrow e^- + e^+ + \gamma$  (1.198\%).  Since 
in our model the $\pi^0$  meson consists of a multitude of electromagnetic 
waves, some of them with energies larger than 2m(e$^\pm$)c$_\star^2$, it seems 
that pair production takes place within the $\pi^0$ meson, and even more so 
in the higher modes of the $\gamma$-branch where the electrons and positrons 
created by pair production tend to settle on mesons, as e.g. in $\eta \rightarrow \pi^+ + 
\pi^- + \pi^0$ (23.2\%) or in the decay $\eta \rightarrow \pi^+ + \pi^- + 
\gamma$ (4.78\%), where the origin of the pair of charges is more apparent.

   Finally we must explain the reason for which the photon lattice or the 
$\gamma$-branch particles are limited in size to a particular value, as 
the experiments tell. Conventional lattice theory using the periodic 
boundary condition does not limit the size of a crystal, and in fact very 
large crystals exist. If, however, the lattice consists of standing 
electromagnetic waves the size of the lattice is limited by the radiation 
pressure. The lattice will necessarily break up at the latest when the 
outward directed radiation pressure is equal to the inward directed 
elastic force which holds the lattice together, in technical terms when 
the radiation pressure is equal to Young's modulus of the crystal. The 
radiation pressure can be calculated and is a function of the sidelength 
of the lattice. Young's modulus of the lattice can be calculated from 
lattice theory. From the equality of both follows, as shown in [19], that 
the sidelength of the photon lattice or of the $\pi^0$ meson must be on the 
order of 10$^{-13}$ cm or about the size of the nucleon, the principal 
uncertainty being the breaking strength of lattices, for which a 
satisfactory solution is not known. 

\section {The mass of the $\pi^0$ meson}

   So far we have studied the ratios of the masses of the particles. We 
will now determine the mass of the $\pi^0$ meson in order to validate that 
the mass ratios link with the actual masses of the particles. The 
energy of the $\pi^0$ meson is 

\centerline{E(m($\pi^0$)) = 134.9764 MeV = 2.1626\,$\cdot\,10^{-4}$ erg.}
\noindent
For the sum of the energies  of all 
frequencies of the lattice we use 
the equation

\begin{equation}  E_\nu = 
\frac{Nh\nu_0}{4\pi^2}\int\limits_{-\pi}^{\pi}\int\limits_{-\pi}^{\pi}f(\phi_1,\phi_2)\,\,d\phi_1d\phi_2 \,
. 
\end{equation}

  This equation originates from B\&K. N is the number of all lattice 
points. The total energy of the frequencies in a cubic lattice is equal 
to the number of the oscillations times the average of the energy of the 
individual frequencies. In order to arrive at an exact value of Eq.(18) 
we have to use the correct value of the radius of the proton, which is 
r$_p$ = (0.88 $\pm$ 0.015)\,$\cdot$\,10$^{-13}$ cm according to [17], in 
contrast to the old r$_p$ = 0.8\,$\cdot\,10^{-13}$ cm we used in [6]. With 
$\emph{a}$ = 10$^{-16}$ cm it follows that the number of all lattice points 
in the cubic lattice is

\centerline{N = 2.854\,$\cdot$\,10$^9$.}
\noindent
The radius of the $\pi^\pm$ mesons has also been measured [20] and after 
further analysis [21] was found to be 0.83\,$\cdot$\,10$^{-13}$ cm, which 
means that within the uncertainty of the radii we have $r_p$ = $r_\pi$.

  If the oscillations are parallel to an axis, the 
limitation of the group velocity is taken into account, that means if 
$\nu = \nu_0\phi$, and the absolute 
values of the frequencies are taken, then the value of the 
double integral in Eq.(18) is 2$\pi^3$ = 62.01. If we use for 
f($\phi_1,\phi_2$) the modified frequency distribution shown in Fig.\,3 
with the absolute values of the frequencies then it 
turns out that the numerical value of the double integral in Eq.(18) is 
66.9 (radians$^2$) for the corrected (1.1) state. With N = 
2.854\,$\cdot$\,10$^9$ and $\nu_0$ = c$_\star/2\pi\emph{a}$ it follows from 
Eq.(18) that the sum of the energy of the frequencies corrected for the group 
velocity limitation in the case of Eq.(9) is E$_{corr}(1.1)$ = 
1.418$\cdot$10$^9$ erg, and for the corrected form of Eq.(12) it is 
E$_{corr}$(1.1) = 1.53\,$\cdot$\,$10^9$ erg. That means in the first case 
that the energy is 6.56$\cdot10^{12}$, and in the second case 7.07$\cdot10^{12}$ times larger than E(m($\pi^0$)). This discrepancy is inevitable, 
because 
the basic frequency of the Fourier spectrum after a collision on the order of 
10$^{-23}$ sec duration is 10$^{23}$ sec$^{-1}$, which means, when 
E = h$\nu$, that one basic frequency alone contains an energy of about 
9m($\pi^0$)c$_\star^2$.
 
   To eliminate this discrepancy we use, instead of the simple 
form E = h$\nu$, the complete quantum mechanical energy of a linear 
oscillator as given by Planck

\begin{equation} E = \frac{h\nu}{e^{h\nu/kT} -\,1}\,\,.
\end{equation}
\noindent
This equation was already used by B\&K for the determination of the 
specific heat of cubic crystals or solids. Equation (19) calls into 
question the value of the temperature T in the interior of a particle. We 
determine T empirically with the formula for the internal energy of solids

\begin{equation} u = \frac{R\Theta}{e^{\Theta/T} -1}\,\,,
\end{equation}
\noindent
which is from Sommerfeld [22]. In this equation it is now R = 
2.854\,$\cdot\,10^9$k, where k is Boltzmann's constant, and $\Theta$ is the 
characteristic temperature introduced by Debye [23] for the explanation of 
the specific heat of solids. It is $\Theta = h\nu_m$/k, where $\nu_m$ is a 
maximal frequency. In the case of the oscillations making up the $\pi^0$ 
meson the maximal frequency is $\nu_m = \pi\nu_0$, see Figs.\,3,4, 
therefore 
$\nu_m = 1.5\cdot10^{26}$ sec$^{-1}$, and we find that $\Theta = 
7.2\cdot10^{15}$\,K.
 
  In order to determine T we set the internal energy u equal to 
m$(\pi^0)$c$_\star^2$. It then follows from Eq.(20) that $\Theta$/T = 30.20, or 
T = 2.38\,$\cdot$\,10$^{14}$ K. That means that Planck's formula (19) introduces a 
factor $1/(e^{\Theta/T} - 1\,) \approx 1/e^{30.2} = 1/(1.305\cdot10^{13}$) into 
Eq.(18). In other words, if we determine the temperature T of the particle through 
equation (20), and correct (18) accordingly then we arrive at a sum of the 
oscillation energies of the $\pi^0$ meson which is 1.0866$\cdot10^{-4}$\,erg = 67.82 MeV in the case of the corrected 
Eq.(9) and 1.172$\cdot10^{-4}$\,erg in the case of the corrected Eq.(12). That 
means that the sum of the energies of the oscillations parallel to an 
axis is 0.502E(m($\pi^0$)), and for the two-dimensional 
oscillations the sum of the energies in the corrected frequencies is 
0.542E(m($\pi^0$)). We have to double this amount because standing waves 
consist of two waves traveling in opposite direction with the same absolute 
value of the frequency. The sum of the energy of the lattice oscillations in the 
$\pi^0$ meson is 
therefore
 
\centerline{E$_\nu$ = 2.1732$\cdot10^{-4}$\,erg = 135.64 MeV = 
1.005E(m($\pi^0$))(exp),}
\noindent
in the case 
that the oscillations are parallel to the $\phi_1$ axis, or 2.345$\cdot10^{-4}$\,erg = 1.084E(m($\pi^0$)) in the case of the two-dimensional oscillations Eq.(12). The energy in the mass of the $\pi^0$ meson and the energy 
in the sum of the lattice oscillations agree fairly well, considering the 
uncertainties of the parameters involved. The energy contained in the 
axial oscillations (Eq.9) matches the measured energy in the $\pi^0$ meson 
much better than the energy contained in the two-dimensional oscillations 
(Eq.12). The calculations therefore favor the interpretation that the oscillations 
in the $\pi^0$ mesons are purely axial. 

   To summarize: We find that the energy in the rest mass of the $\pi^0$ 
meson and the other particles of the $\gamma$-branch are correctly given 
by the sum of the energy of the standing electromagnetic waves, if the 
energy of the oscillations is determined by Planck's formula for the 
energy of a linear oscillator. The $\pi^0$ meson is like an adiabatic, 
cubic black body filled with standing electromagnetic waves. We know from 
Bose's work [24] that Planck's formula applies to a photon gas as well.
For all $\gamma$-branch particles we have found a simple mode of standing 
waves in a cubic lattice. For the explanation of the mesons and baryons of 
the $\gamma$-branch $\emph{we use only photons, nothing else}$. A rather 
conservative explanation of the $\pi^0$ meson, and the $\gamma$-branch 
particles. It is worth noting that in the $\gamma$-branch of our model 
there is a continuous line leading from the creation of a particle out of 
photons or electromagnetic waves through the lifetime of the particle as 
standing electromagnetic waves to the decay products which are 
electromagnetic waves as well.

   From the frequency distributions of the standing waves in the lattice 
follow the ratios of the masses of the particles which obey the integer 
multiple rule. It is important to note that in this theory the ratios of 
the masses of the $\gamma$-branch particles to the mass of the $\pi^0$ 
meson $\emph{do not depend}$ on the sidelength of the lattice, and the 
distance between the lattice points, neither do they depend on the 
strength of the force between the lattice points nor on the mass of the 
lattice points. The mass ratios are determined only by the spectra of the 
frequencies of the standing waves in the lattice. Since the equation 
determining the frequency of the standing waves is quadratic it follows 
\emph{automatically} that for each positive frequency there is also a negative 
frequency of the same absolute value, that means that for each particle 
there exists also an antiparticle.

\section {The neutrino branch particles}

The masses of the neutrino branch, the $\pi^\pm$, K$^{\pm,0}$, n, 
D$^{\pm,0}$ and D$^\pm_S$ particles, are integer multiples of the mass of 
the $\pi^\pm$ mesons times a factor $0.85\,\pm\,0.02$ as we stated before. 
We have explained the integer multiple rule of the $\gamma$-branch 
particles with different modes and superpositions of standing 
electromagnetic waves in a cubic nuclear lattice. For the explanation of 
the particles of the neutrino branch we follow a similar path. As we have 
done in [25] we assume, as appears to be quite natural, that the $\pi^\pm$ 
mesons and the other particles of the $\nu$-branch $\emph{consist of the 
same particles into which they decay}$, that means of neutrinos and 
electrons or positrons. Since the particles of the $\nu$-branch decay 
through weak decays, we assume, as appears likewise to be natural, that 
\emph{the weak nuclear force holds the particles of the $\nu$-branch together}. 
Since the range of the weak interaction is only about a thousandth of the 
diameter of the particles, the weak force can hold particles together only 
if the particles have a lattice structure, just as macroscopic crystals 
are held together by microscopic forces between atoms. We will, therefore, 
investigate the energy which is contained in the oscillations of a cubic 
lattice consisting of electron and muon neutrinos and in the rest masses of 
the neutrinos.

   Since we will investigate the 
oscillations of a cubic lattice consisting of muon and electron neutrinos 
it is necessary to outline the basic aspects of diatomic lattice 
oscillations. In $\emph{diatomic}$ lattices the lattice points have 
alternately the masses m and M. The classic example of a diatomic lattice 
is the salt crystal with the masses of the Na and Cl atoms in the lattice 
points. The theory of diatomic lattice oscillations was developed by Born 
and v.Karman [9]. They first discussed a diatomic chain. The equation of 
motions in the chain are according to Eq.(22) of B\&K

\begin{equation} m\ddot{u}_{2n} = \alpha(u_{2n+1} + u_{2n-1} - 2u_{2n})\, 
,\end{equation}

\begin{equation} M\ddot{u}_{2n+1} = \alpha(u_{2n+2} + u_{2n} - 
2u_{2n+1}) \,, 
\end{equation}
\noindent    
where the u$_n$ are the displacements, n an integer number and $\alpha$ a 
constant characterizing the force between the particles. As with any 
spring the restoring forces in (21,22) increase with increasing distance 
between the particles. Eqs.(21,22) are solved with

\begin{equation}u_{2n} = Ae^{i(\omega\,t\,+\,2n\phi)} ,\end{equation} 

\begin{equation}u_{2n+1} = Be^{i(\omega\,t\,+\,(2n+1)\phi)}\,,\end{equation} 

\noindent
where A and B are constants and $\phi$ is given by $\phi = 
2\pi\emph{a}/\lambda$ as in (8). $\emph{a}$ is the lattice constant as 
before and $\lambda$ the wavelength, $\lambda$ = n$\emph{a}$. The 
solutions of Eqs.(23,24) are obviously periodic in time and space and 
describe again standing waves. Using (23,24) to solve (21,22) leads to a 
secular equation from which according to Eq.(24) of B\&K the frequencies of 
the oscillations of the chain follow from 

\begin{equation}4\pi^2\nu^2_\pm  = \alpha/Mm\cdot((M+m) \pm\sqrt{(M-m)^2 + 
4mMcos^2\phi}\,)\,.\end{equation}    

Longitudinal and transverse waves are distinguished by the minus or plus 
sign in front of the square root in (25). For a 
similar equation for the plane waves in an isotropic three-dimensional 
lattice see [25]. The equations of motion for the oscillations in a 
three-dimensional diatomic lattice have been developed by Thirring [26].

\section{The masses of the $\nu$-branch particles}

The particles of the neutrino branch decay primarily by weak decays, see 
Table 4. The characteristic case are the $\pi^\pm$ mesons which decay via 
e.g. $\pi^+ \rightarrow\mu^+ + \nu_\mu$ (99.988\%) followed by  $\mu^+ 
\rightarrow e^+ + \bar{\nu}_\mu + \nu_e$ ($\approx$ 100\%). Only $\mu$ 
mesons, which decay into neutrinos, and neutrinos 
result from the decay of the $\pi^\pm$ mesons, but for e$^\pm$ which 
conserve charge. If the particles consist of the particles into which they 
decay, then the $\pi^\pm$ mesons and the other particles of the neutrino 
branch are made of neutrinos and e$^\pm$. The neutrinos must be held 
together in some form, otherwise the particles could not exist over a 
finite lifetime, say 10$^{-10}$ sec. Since neutrinos interact through the 
weak force which has a range of 10$^{-16}$ cm according to p.25 of [16], 
and since the size of the nucleon is on the order of 10$^{-13}$ cm, we suggest that 
the $\nu$-branch particles are held together in a cubic lattice. A cubic 
lattice is held together by central forces, which are the most simple 
forces to consider. It is not known with certainty that neutrinos actually 
have a rest mass as was originally suggested by Bethe [27] and Bahcall 
[28] and what the values of m($\nu_e$) and m($\nu_\mu$) are. However, the 
results of the Super-Kamiokande [29] and the Sudbury [30] experiments 
indicate that the neutrinos have rest masses. The neutrino lattice must be 
diatomic, meaning that the lattice points have alternately larger 
(m($\nu_\mu$)) and smaller (m($\nu_e$)) masses. We will retain the 
traditional term diatomic. The lattice we consider is shown in Fig.\,5. 
Since the neutrinos have spin 1/2 this is a four-Fermion lattice as is 
required for Fermi's explanation of the $\beta$-decay. The first 
investigation of cubic Fermion lattices in context with the elementary 
particles was made by Wilson [7]. The entire neutrino lattice is 
electrically neutral. Since we 
do not know the structure of the electron we cannot consider lattices with a charge.

	\begin{figure}[h]
	\hspace{3cm}
	\includegraphics{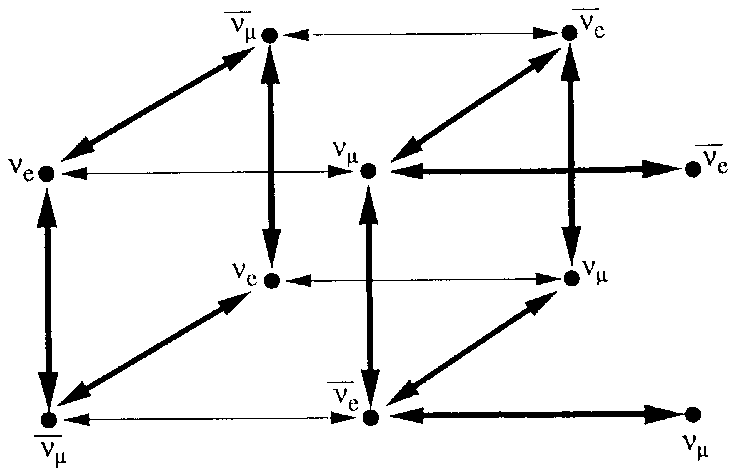}
	\vspace{-0.2cm}
	\begin{quote}
Fig.\,5: The neutrino lattice. Bold lines mark the forces between 
neutrinos and antineutrinos. Thin lines mark the forces between either 
neutrinos only, or antineutrinos only.
	\end{quote}
	\end{figure}

A neutrino lattice takes care of the continuum of frequencies which must, 
according to Fourier analysis, be present after the high energy collision 
which created the particle. We will, for the sake of simplicity, first set 
the sidelength of the lattice at 10$^{-13}$ cm that means approximately 
equal to the size of the nucleon. The lattice then contains 10$^9$ lattice 
points, since the lattice constant $\emph{a}$ is on the order of 10$^{-16}$ cm. 
The sidelength of the lattice does not enter the equations, e.g. Eq.(25), 
for the frequencies of the lattice oscillations. The calculation of the 
ratios of the masses or of the energy of the lattice oscillations is 
consequently independent of the size of the lattice, as was the case with 
the $\gamma$-branch. However the size of the lattice can be explained with 
the pressure which the lattice oscillations exert on a crossection of the 
lattice. The pressure cannot exceed Young's modulus of the lattice. We require that the lattice is 
isotropic.

	\begin{figure}[h]
	\hspace{1.5cm}
	\includegraphics{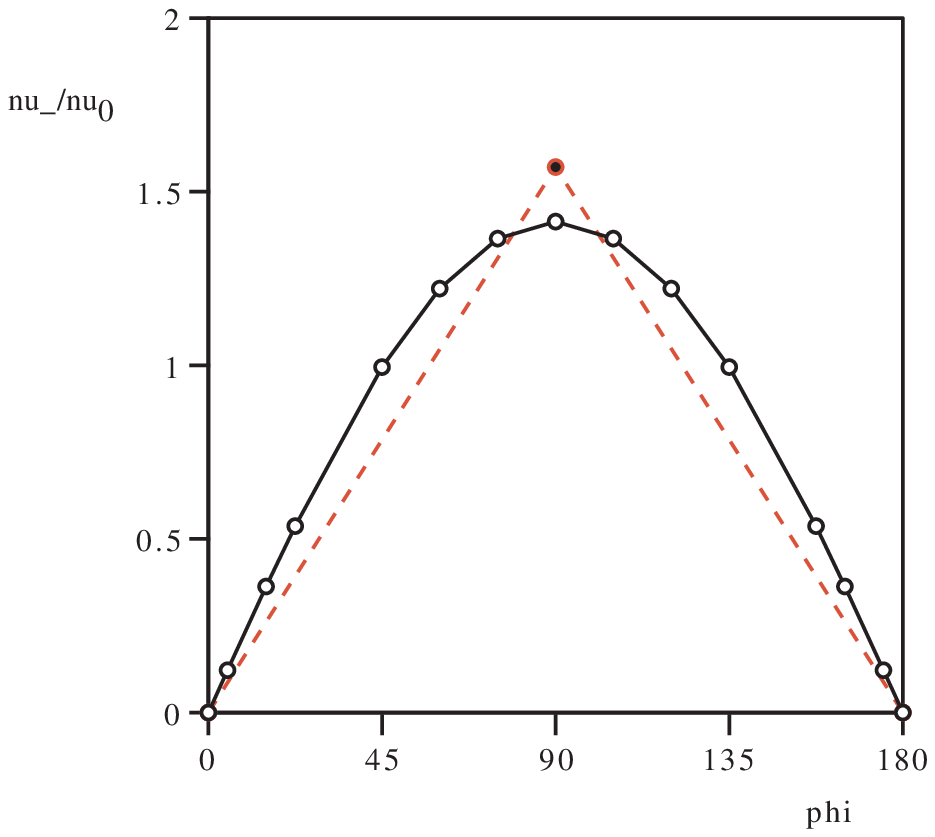}
	\vspace{-.5cm}
	\begin{quote}
Fig.\,6: The frequency distribution $\nu_-/\nu_0$ of the basic diatomic mode 
according to Eq.(25) with M/m = 50. The dashed line shows the distribution 
of the frequencies corrected for the group velocity limitation.
	\end{quote}
	\end{figure}

   From the frequency distribution of the axial diatomic oscillations 
   (Eq.\,25), shown in Fig.\,6, follows the group velocity 
   $\mathrm{d}\omega/\mathrm{dk} 
= 2\pi\emph{a}\,\,d\nu/d\phi$\, at each point $\phi$. With $\nu = 
\nu_0f(\phi)$ and $\nu_0$ = c$_\star/2\pi\emph{a}$ from Eq.(14) we find

\begin{equation} c_g = d\omega/dk = 
\emph{a}\sqrt{\alpha/M}\cdot d\,f(\phi)/d\phi\,.
\end{equation}
\noindent	
In order to determine the value of d$\omega/dk$ we have to know the value 
of $\sqrt{\alpha/M}$. From Eq.(13) for $\alpha$ follows that $\alpha = 
\emph{a}\,c_{44}$ in the isotropic case with central forces. The group 
velocity is therefore

\begin{equation} c_g = \sqrt{a^3c_{44}/M}\cdot df/d\phi\,.
\end{equation}

\noindent
We now set $\emph{a}\sqrt{\alpha/M}$ = c$_\star$ as in Eq.(16), where 
c$_\star$ is the velocity of light. It follows that

\begin{equation} c_g = c_\star\cdot df/d\phi\,,
\end{equation}

\noindent
as it was with the $\gamma$-branch, only that now on account of 
the rest masses of the neutrinos the group velocity must be smaller than 
c$_\star$, so the value of df/d$\phi$ is limited to $<$\,1, but c$_g \cong$\, c$_\star$, 
which is a necessity because the neutrinos in the lattice soon approach 
the velocity of light as we will see. Equation (28) applies regardless 
whether we consider $\nu_+$ or $\nu_-$ in  
Eq.(25). That means that there are no separate transverse 
oscillations with their theoretically higher frequencies.
 
From Eq.(27) follows a formula for the mass of the muon neutrino, it is

\begin{equation} M\,=\,\emph{a}^3c_{44}/c_\star^2\,.
\end{equation}

\noindent
The elasticity constant c$_{44}$ = c$_{11}$/3 can be determined theoretically 
from an exact copy of the determination of c$_{11}$ in Born's lattice 
theory, assuming weak nuclear forces between lattice points at distance 
$\emph{a}$ = 10$^{-16}$ cm, and replacing $e^2$ by $g^2$, where $g^2$ is 
the interaction constant of the weak force. This will be discussed in 
section 8. From c$_{11}$ in Eq.(47) we find that

\begin{equation} M = \mathrm{m}(\nu_\mu) = 0.538\,g^2(n - 1)/3\emph{a}c_\star^2\,,
\end{equation}
\noindent
and with $\emph{a} = 10^{-16}$ cm, $g^2 = 2.946\cdot 10^{-17}$\,erg$\cdot$cm 
(Eq.43) and (n $-$ 1) = 2.187$\,\cdot\,10^{-12}$ from Eq.(44) we have

\begin{equation} \mathrm{m}(\nu_\mu) = 1.28\cdot 10^{-34}\,\mathrm{gr} = 
72.1\cdot 10^{-3}\,\mathrm{eV/c_\star^2}\,.
\end{equation}

The rest mass of the 
muon neutrino is 72 milli-electron-Volt/c$_\star^2$, within the accuracy of the parameters $g^2$, $\emph{a}$\ and (n $-$ 1). We note that (n $-$ 1) 
depends on the compressibility $\kappa$ for which values are given in [35] which 
differ by 33\%. We will later on determine the rest mass of the muon neutrino from the 
difference m($\pi^\pm$)\,$-$\,m($\mu^\pm$), which leads to m($\nu_\mu$) = 
47.5 meV/c$_\star^2$ (Eq.35). We use, henceforth, only this value because 
it depends on a single parameter, the number of muon neutrinos in the 
lattice.

  It can be verified easily that 
the mass m($\nu_\mu$) = 47.5 meV/c$_\star^2$ we have found makes sense. The energy of the 
rest mass of the $\pi^\pm$ mesons is 139 MeV, and we have N/4 = 0.7135$\cdot10^9$ 
muon neutrinos and the same number of anti-muon neutrinos. It 
then follows that the energy in the rest masses of all muon and 
anti-muon neutrinos is 
67.8 MeV, that is 48.5\% of the energy of the rest mass of the $\pi^\pm$ mesons. A small 
part of m($\pi^\pm$)c$_\star^2$ goes, as we will see, into the electron neutrino 
masses, the rest goes 
into the lattice oscillations.

   We can now determine the rest mass of the $\pi^\pm$ mesons from the sum 
of the oscillation energies and the sum of the rest masses of the 
neutrinos. For the sum of the energies of the frequencies we use Eq.(18) with the 
same N and $\nu_0$ we used for the $\gamma$-branch. For the double integral 
in (18) of the 
corrected axial 
diatomic frequencies we find the value 
$\pi^3$ as can be easily derived from the plot of the corrected frequencies 
in Fig.\,6. The value of the 
double integral in Eq.(18) for the axial diatomic frequencies $\nu = 
\nu_0\phi$ is 1/2 of 
the value 2$\pi^3$ of the same integral in the case of axial monatomic 
frequencies, because in the latter case the increase of the corrected 
frequencies continues to $\phi_1$ = $\pi$, whereas in the diatomic case the 
increase of the corrected frequencies ends at $\pi$/2, see Fig.\,6.  We consider c$_g$ 
to be so close to c$_\star$ that it does not 
change the value of the double integral significantly. It can be 
calculated that the time average of 
the velocity of the electron neutrinos in the $\pi^\pm$ mesons is 
$\bar{v}$ = 0.99993c$_\star$, if m($\nu_e$) = 0.55 meV/c$_\star^2$ as in Eq.(36). 
Consequently we find that the sum of the energies of the corrected diatomic neutrino 
frequencies is 0.5433$\cdot$10$^{-4}$\,erg = 33.91 MeV. We double this 
amount because we deal with standing waves and find that the energy of the neutrino oscillations 
is 67.82 MeV. Adding to 
that the sum of the energy of the rest masses of the neutrinos N/2$\cdot$(47.5 + 0.55)meV = 68.57 MeV we obtain a value 
for
\medskip
\centerline{ m($\pi^\pm)$c$_\star^2$(theor) = 136.39 MeV =  
0.977m($\pi^\pm)$c$_\star^2$(exp) = 0.977$\cdot$139.57 MeV.}
\smallskip
\noindent
To this theoretical value of the mass we must add m($e^\pm$)c$_\star^2$ in order to obtain 
the mass of the $\pi^\pm$ mesons plus or minus some possible binding energy.
We must keep in mind that the relation between the theoretical and 
experimental m($\pi^\pm$) still depends on the accuracy of the values of N and $\nu_0$ and 
on a small correction due to the possible presence of surface waves on the lattice.

   We have explained the $\pi^\pm$ mesons with neutrinos. This agrees 
with what we learn from the decay of the $\pi^\pm$ mesons, 
99.9877\% of which decay with the emission of a muon neutrino and a $\mu$ 
meson, which in turn 
consists of neutrinos as well, as we will see in the next section. The 
antiparticle of the $\pi^+$ meson is the particle in which all frequencies 
of the neutrino lattice oscillations have been replaced by frequencies with 
the opposite sign, all neutrinos replaced by their antiparticles and the 
positive charge replaced by the negative charge. Assuming that the 
antineutrinos have the same rest mass as the neutrinos it follows that the 
antiparticle of the $\pi^+$ meson has the same mass as $\pi^+$ but 
opposite charge, i.e. is the $\pi^-$ meson. 
 
   The primary decay of the K$^\pm$ mesons, say, K$^+ \rightarrow \mu^+ + \nu_\mu$  
(63.5\%), leads to the same end products as the $\pi^\pm$ meson decay 
$\pi^+ \rightarrow \mu^+ + \nu_\mu$. From this and the composition of the $\mu$ mesons we learn 
that the K mesons must, at least partially, be made of the same four neutrino types 
as the $\pi^\pm$ 
mesons namely of muon neutrinos, 
anti-muon neutrinos, electron neutrinos and anti-electron neutrinos and 
their oscillation energies. However the K$^\pm$ mesons cannot be the (2.2) mode of the lattice 
oscillations of the 
$\pi^\pm$ mesons, because the (2.2) mode of the neutrino 
lattice oscillations 
has an energy of 4E$_{\nu}$($\pi^\pm$) + N/2$\cdot$(m($\nu_\mu$) + 
m($\nu_e$))\,c$_\star^2$ = (271.3 + 68.57)MeV = 340 MeV. The 340 MeV characterize 
the (2.2) mode of the $\pi^\pm$ mesons, which fails 
m(K$^\pm$)c$_\star^2$ = 493.7 MeV by a wide margin.

   Anyway, the concept of the K$^\pm$ mesons being solely a higher mode of the 
$\pi^\pm$ mesons contradicts our point that the particles consist of the 
particles into which they decay. The decays K$^\pm$ $\rightarrow$ 
$\pi^\pm$ + $\pi^0$ (21.16\%), as well as K$^+$ $\rightarrow$ $\pi^0 + 
e^+ + \nu_e$ (4.82\%) and K$^+$ $\rightarrow$ $\pi^0 + \mu^+ + 
\nu_\mu$ (3.18\%) make up 29.16\% of the K$^\pm$ meson decays. A $\pi^0$ meson figures 
in each of these decays. If we add the energy in the rest mass of a $\pi^0$ 
meson m($\pi^0$)c$_\star^2$ = 134.97 MeV to the 340 MeV in the (2.2) mode of the $\pi^\pm$ 
meson then we arrive at an energy of 475 MeV, which is 96.2\% of 
m(K$^\pm$)c$_\star^2$. Therefore we conclude that the K$^\pm$ mesons consist of the 
(2.2) mode of the $\pi^\pm$ mesons $\emph{plus}$ a $\pi^0$ meson. Then it 
is natural that $\mu$ mesons from the decay of the (2.2) mode of the 
$\pi^\pm$ mesons in 
the K$^\pm$ mesons as well as $\pi^0$ mesons from the $\pi^0$ component in the 
K$^\pm$ mesons appear as decay 
products of the K$^\pm$ meson. It may be coincidental but is interesting 
that the ratio 0.273 of the energy in the $\pi^0$ meson to the energy in the K$^\pm$ 
mesons closely resembles the 29.16\% of the decays 
in which the $\pi^0$ meson is one of the decay products.

   We obtain the K$^0$ meson if we superpose onto the (2.2) mode of the $\pi^\pm$ mesons instead of 
a $\pi^0$ meson a basic mode of the $\pi^\pm$ mesons with a charge opposite 
to the charge of the (2.2) mode of the $\pi^\pm$ meson. The K$^0$ mesons, or 
the state (2.2)$\pi^\pm$ + $\pi^\mp$,  
is made of neutrinos only without a photon component, because the (2.2) mode 
of $\pi^\pm$ as well as the basic mode $\pi^\mp$ consist of neutrinos only. Since the mass of a $\pi^\pm$ meson is by 4.59 MeV/c$_{\star}^2$ 
larger than the mass of a $\pi^0$ meson the mass of K$^0$ should be larger than 
m(K$^\pm$), and indeed m(K$^0$)\,$-$\,m(K$^\pm$) = 3.995 MeV/c$_\star^2$ 
according to [2]. The 
decay K$^0_S \rightarrow \pi^+ + \pi^-$ (68.6\%) creates directly the 
$\pi^+$ and $\pi^-$ mesons which are part of the (2.2)$\pi^\pm$ + $\pi^\mp$ 
structure of K$^0$ we have 
suggested. The decay K$^0_S \rightarrow \pi^0 + \pi^0$ (31.4\%) apparently 
originates from the 2$\gamma$ branch of electron positron annihilation. 
Both decays account for 100\% of the decays of K$^0_S$. 
The decay K$^0_L \rightarrow 3\pi^0$ (21.1\%) apparently comes from the 
3$\gamma$ branch of electron positron annihilation. The two decays of 
K$^0_ L$ called K$^0_{\mu3}$ and K$^0_{e3}$ which together make up 65.95\% of 
the K$^0_{L}$ decays apparently originate from the decay of either of the 
(2.2) mode of the $\pi^\pm$ mesons or the basic mode of $\pi^\mp$ in the 
K$^0$ structure, either into $\pi^\pm$ + $\mu^\mp$ + $\nu_\mu$  
 or into $\pi^\pm$ + $e^\mp$ + $\nu_e$. Our rule 
that the particles consist of the particles into which they decay also 
holds for the K$^0$ meson.   

   The neutron is the superposition of a K$^+$ and a K$^-$ meson or of two 
K$^0$ mesons. The neutrino oscillations in the neutron must be 
coupled  in order to have spin 1/2, just as the $\Lambda$ 
baryon with spin 1/2 is a superposition of two $\eta$ mesons. With 
m(K$^\pm$)(theor) = 475 MeV/c$^2_\star$ from above it follows that m(n)(theor) = 
2m(K$^\pm)$(theor) = 950 MeV/c$^2_\star$ = 1.011m(n)(exp). The D$^\pm$ mesons are the 
superposition of a proton and a neutron of opposite spin and hence have no 
spin, whereas the 
superposition of a proton and a neutron with the same spin creates the 
deuteron with spin 1. In this case the proton and neutron interact with the strong 
force which will be discussed in the last section. The D$_S^\pm$ mesons seem 
to be the superposition of a $\pi^0$ meson on the D$^\pm$ mesons. 

   The proton does not decay and does not tell which particles it is 
made of. However we learn about the structure of the proton through the decay 
of the neutron n $\rightarrow$ p + e$^- + \bar{\nu}_e$ (100\%). 
One single anti-electron neutrino is emitted when the neutron decays and 1.3 
MeV are released. But there 
is no place for a permanent vacancy of a single missing neutrino and for a small 
amount of permanently missing oscillation energy in a nuclear lattice. As it appears 
the entire anti-electron neutrino type is removed from the structure of 
the neutron in the neutron decay and converted to available energy. This 
process will be discussed again in the following 
section on the $\mu$ meson. On the other hand, it seems to be certain that 
the proton contains a cubic lattice consisting of four muon and four 
anti-muon neutrinos in each lattice cell. This lattice exists already in the 
superposition of a K$^+$ and a K$^-$ or of two K$^0$ mesons which make up the neutron. The 
muon-anti-muon neutrino lattice is not affected by the decay of the 
neutron, no muon neutrino of either type is emitted. The neutron from the 
K$^+$ and K$^-$ mesons also 
contains the superposition of two basic photon lattice oscillations, which are 
not affected by the neutron decay either, otherwise the energy released could not 
be only 1.3 MeV. So the proton contains a cubic muon neutrino-anti-muon 
neutrino lattice and the superposition of two $\pi^0$ mesons. Whether or 
not N/2 electron neutrinos are part of the proton structure awaits 
clarification of the structure of the electron.

   The 
average 
factor 0.85 $\pm$ 0.025 in the ratios of the particles of the $\nu$-branch 
to the $\pi^\pm$ mesons is a consequence of the neutrino rest masses. They 
make it impossible that the ratios of the particle masses are integer multiples because the 
particles consist of the energy in the neutrino oscillations plus the 
neutrino rest masses which are independent of the order of the lattice 
oscillations. Since the contribution in percent of the neutrino rest 
masses to the $\nu$-branch particle masses decreases with increased particle 
mass the factor in front of the mass ratios of the $\nu$-branch particles 
must decrease with increased particle mass.  

   To summarize: We have found that the $\pi^\pm$ mesons can be described 
by the oscillations in a cubic lattice consisting of muon neutrinos and 
electron neutrinos and their antiparticles. The energy in the $\pi^\pm$ 
mesons is the sum of the oscillation energies plus the energy in the rest 
masses of the neutrinos. For the explanation of the K$^\pm$ mesons, as well as 
for the explanation of the neutron and the D mesons it is 
necessary to have also a $\pi^0$ meson present in the particles. When 
the $\pi^0$ meson is incorporated in the K$^\pm$ meson the decay modes of the 
K$^\pm$ 
meson appear to be natural. For the explanation of the neutrino branch 
particles we need neutrinos and photons.

\section{The mass of the $\mu^\pm$ mesons}

Surprisingly one can also explain the mass of the $\mu^\pm$ mesons with the 
standing wave model as we have shown in [31]. The $\mu$ mesons are part of 
the lepton family which is distinguished from the mesons and baryons not 
so much by their mass as the name lepton implies, actually the mass of the 
$\tau$ meson is about twice the mass of the proton, but rather by the 
absence of strong interaction with the mesons and baryons. The masses of 
the leptons are not explained by the standard model of the particles.

   The mass of the $\mu$ mesons is m($\mu^\pm$) = 105.658389 $\pm$ 
3.4\,$\cdot$\,$10^{-5}$ MeV/c$_\star^2$, according to the Particle Physics 
Summary [2]. The $\mu$ mesons are ``stable", their lifetime is $\tau(\mu^\pm) 
= 2.19703\cdot10^{-6} \pm 4\cdot10^{-11}$\,sec, about a hundred times 
longer than the lifetime of the $\pi^\pm$ mesons, that means longer than the 
lifetime of any other elementary particle, but for the electrons, protons 
and neutrons.

   Comparing the mass of the $\mu$ mesons to the mass of the $\pi^\pm$ mesons 
m($\pi^\pm$) = 139.56995 MeV/c$_\star^2$ we find that 
m($\mu^\pm$)/m($\pi^\pm$) = 0.757028 $\approx$ 3/4 or that m($\pi^\pm)\,-$ 
m($\mu^\pm$) = 33.9116 MeV/c$_\star^2$ = 0.24297m($\pi^\pm$) or approximately 
1/4\,$\cdot$\,m$(\pi^\pm$). The mass of the electron is approximately 1/206 of 
the mass of the muon, the contribution of m(e$^\pm$) to m($\mu^\pm$) will therefore 
be neglected in the following. We assume, as we have done before and as 
appears to be natural, that the particles, including the muons, 
$\emph{consist of the particles into}$ \emph{which they decay}. The $\mu^+$ meson 
decays via $\mu^+ \rightarrow e^+ + \bar{\nu}_\mu + \nu_e$ ($\approx$ 
100\%). The muons are apparently composed of some of the neutrinos and their 
oscillations which 
are present in the cubic neutrino lattice of the $\pi^\pm$ mesons 
according to our standing wave model. The $\pi^+$ meson decays via $\pi^+ 
\rightarrow \mu^+ + \nu_\mu$ or the conjugate particles in the decay of 
the $\pi^-$ meson, with 99.988\% of the $\pi^\pm$ decays in this form. The 
energy m($\pi^\pm$)c$_{\star}^2\,-\,$m($\mu^\pm$)c$_{\star}^2$ $\approx 
1/4\cdot$m$(\pi^\pm)$c$_{\star}^2$ is lost when a $\mu^+$ meson and one muon 
neutrino $\nu_\mu$ are 
emitted by the $\pi^+$ mesons. The rest of the energy in the rest mass 
of the $\pi^\pm$ mesons passes to the rest mass of the $\mu^\pm$ mesons.

   In the standing wave model of the particles of the neutrino branch the 
$\pi^\pm$ mesons are composed of a cubic lattice consisting of N/4 = 
0.7135\,$\cdot\,10^9$ muon neutrinos $\nu_\mu$ and the same number of 
anti-muon neutrinos $\bar{\nu}_\mu$, (m($\nu_\mu$) = m($\bar{\nu}_\mu$)), 
as well as of N/4 electron neutrinos $\nu_e$ and the same number of 
anti-electron neutrinos $\bar{\nu}_e$, (m($\nu_e$) = m($\bar{\nu}_e$)), 
plus the oscillation energy of these neutrinos. We say that the mass of 
a single muon neutrino should be 47.5 milli-eV/c$_\star^2$ (Eq.(35)), and 
the mass of a single electron neutrino should be 0.55 meV/c$_\star^2$ 
according to Eq.(36). With 
these values of N and of the neutrino masses we find that:
\smallskip

(a) The difference of the rest masses of the $\mu^\pm$ and $\pi^\pm$ 
mesons is nearly equal to the sum of the rest masses of all muon, 
respectively anti-muon, neutrinos in the $\pi^\pm$ mesons.
\bigskip

\noindent
    m($\pi^\pm)\,-\,$m$(\mu^\pm$) = 33.912 MeV/$c_\star^2$ \quad versus 
\quad N/4$\cdot$m($\nu_\mu$) = 33.89 MeV/c$_\star^2$\,.
\bigskip

(b) The energy in the oscillations of all neutrinos in the      $\pi^\pm$ 
mesons is nearly the same as the energy in the            oscillations of 
all $\bar{\nu}_\mu$, respectively $\nu_\mu$,        and the $\nu_e$ and 
$\bar{\nu}_e$ neutrinos in the $\mu^\pm$            mesons.
\begin{equation}    E_{\nu}(\pi^\pm) = \mathrm{m}(\pi^\pm)c_\star^2 - 
N/2\cdot[\mathrm{m}(\nu_\mu) + 
\mathrm{}m(\nu_e)]c_\star^2 = 71.0 \mathrm{MeV} \end{equation}
\quad versus 
\begin{equation} E_{\nu}(\mu^\pm) = \mathrm{m}(\mu^\pm)c_\star^2 - 
N/4\cdot \mathrm{m}(\bar{\nu}_\mu)c_\star^2 - N/2\cdot\mathrm{m}(\nu_e)c_\star^2 = 
70.98 \mathrm{MeV}\,.\end{equation}

Both statements are, of course, valid only within the accuracy with which 
the number N of all neutrinos in the $\pi^\pm$ lattice is known, as well 
as within the accuracy with which the masses m($\nu_\mu$) = 
m($\bar{\nu}_\mu$) and m($\nu_e$) = m($\bar{\nu}_e$) 
have been determined. It cannot be expected that this accuracy is better 
than a few percent, considering in particular the uncertainty of the 
lattice constant. If E$_{\nu}(\pi^\pm)$ = E$_{_\nu}(\mu^\pm)$ then it 
follows from the difference of Eq.(32) and Eq.(33) that 
\begin {equation} \mathrm{m}(\pi^\pm) - 
\mathrm{m}(\mu^\pm) = N/4\cdot\mathrm{m}(\nu_\mu)\,. 
\end{equation}

   We note that according to (a) and Eq.(34) the difference of the rest masses 
   m($\pi^\pm)\,-$\, 
m($\mu^\pm)$ provides an independent check of the value of the rest mass 
of the muon neutrino. With N/4 = 0.7135\,$\cdot\,10^9$ and 
m($\pi^\pm)\,-\, 
\mathrm{m}(\mu^\pm)$ = 33.912 MeV/c$_\star^2$ it follows that the rest mass of the 
muon neutrino should be 
\begin{equation}\mathrm{m}(\nu_\mu) = 47.5\,\mathrm{meV/c}_\star^2\,, \end{equation}
 whereas we 
found m($\nu_\mu)$ = 72 meV/c$_\star^2$ in Eq.(31) from an entirely 
theoretical determination. We should note that in the $\pi^\pm$ decays 
only \emph{one single} muon neutrino is emitted, not N/4 of them. The energy in 
the rest masses of the N/4 $-$ 1 other muon neutrinos is used to supply the 
kinetic  
energy in the momentum of the emitted muon neutrino (pc$_\star$ = 30\,MeV) and in 
the momentum of the emitted $\mu$ meson (pc$_\star$ = 4\,MeV).

 If the same principle that applies to the decay 
of the $\pi^\pm$ mesons, namely that in the decay an entire neutrino type 
is removed from the neutrino lattice, also applies to the decay of the 
neutron $n\,\rightarrow\,p + e^- + \bar{\nu}_e$, then the mass of the 
anti-electron neutrino can be determined from the known difference $\Delta$ =  
m(n)\,$-$\,m(p) = 1.29332 MeV/c$_\star^2$. Nearly one half of $\Delta$ comes 
from the energy lost by the emission of the electron, whose mass is 
0.510999 MeV/c$_\star^2$. N/2 anti-electron neutrinos are in the neutron, 
which is the superposition of two K$^\pm$ or K$^0$ mesons each containing N/4 neutrinos 
of the four neutrino types. From $\Delta$ $-$ 
m($e^-$)c$_\star^2$ = 
0.782321 MeV follows, after division by N/2, that 
\begin {equation} \mathrm{m}(\nu_e) = 0.55\,\mathrm{meV/c}_\star^2\,,\end{equation}
assuming m($\nu_e$) = m($\bar{\nu}_e$). From the magnitude of $\Delta$ 
follows also that the oscillation energy in the neutron must be preserved 
in the neutron decay, as it is with the oscillation energy in the 
$\mu^\pm$ decay according to (b).

   We have still to account for one type of electron neutrino which, 
acording to our model of the $\pi^\pm$ mesons, is part of the 
neutrino lattice of the $\pi^\pm$ mesons but does not appear in the decay of, say, the $\pi^+$ 
meson $\pi^+\,\rightarrow\,\mu^+$ + $\nu_\mu$ followed by 
$\mu^+\,\rightarrow$ e$^+$ + $\bar{\nu}_\mu$ + $\nu_e$. The missing 
anti-electron neutrino $\bar{\nu}_e$ in the $\pi^+$ decay sequence must go 
with the emitted positron. Whether or not this interpretation is correct 
can be decided only after the explanation of the structure of the electrons or 
positrons. 

   Since according to the statement (a) the decay of the $\pi^\pm$ meson seems to mean the removal of all 
$\nu_\mu$, respectively, all $\bar{\nu}_\mu$ neutrinos from the neutrino 
lattice of the $\pi^\pm$ mesons, the $\mu$ mesons should contain the 
remaining neutrinos of the original cubic lattice, that means N/4 
anti-muon neutrinos $\bar{\nu}_\mu$, respectively, N/4 muon neutrinos 
$\nu_\mu$, plus N/4 electron neutrinos $\nu_e$ as well as N/4 
anti-electron neutrinos $\bar{\nu}_e$.           If we use the value of 
m($\nu_\mu$) = 47.5meV/c$_\star^2$ from Eq.(35) we find that 
$E_\nu(\mu^\pm)$/$E_\nu(\pi^\pm)$ = 0.9997, as in 
statement (b). We will show in the following why the energies in the 
frequencies of $\pi^\pm$ and $\mu^\pm$ 
mesons are the same.

   The neutrinos in the body of the $\mu$ mesons must oscillate because the 
collision $e^+ + e^- \rightarrow \mu^+ + \mu^-$ tells that a continuum of 
frequencies must be present in the $\mu$ mesons, if Fourier analysis 
holds.  The 
energy of the oscillations of the neutrinos in the $\mu$ meson lattice is 
the sum of the energies of a longitudinal oscillation in a 
diatomic lattice consisting of N/4\,$\cdot$\,$\bar{\nu}_\mu$ neutrinos and 
N/4\,$\cdot$\,$\nu_e$ neutrinos, part of the remains of the diatomic neutrino lattice of 
the $\pi^\pm$ mesons, plus the energy of the diatomic oscillations of 
N/4\,$\cdot$\,$\bar{\nu}_\mu$ and N/4\,$\cdot$\,$\bar{\nu}_e$ neutrinos which were likewise in the 
neutrino lattice of the $\pi^\pm$ mesons. The latter oscillations are 
likely to be perpendicular to the first mentioned $\bar{\nu}_\mu$-$\nu_e$ oscillations 
because the $\bar{\nu}_e$-$\nu_e$ neutrino pairs are oriented 
perpendicular to the $\bar{\nu}_\mu$-$\bar{\nu}_e$ neutrino pairs in the 
original cubic lattice of the $\pi^\pm$ mesons, see Fig.\,5.

   The energy in the frequencies of the lattice is given by

\begin{equation}E_\nu=\frac{Nh\nu_0}{(2\pi)^2}\int\limits_{-\pi}^{\pi}\int\limits_{-\pi}^{\pi}\,f(\phi_1,\phi_2)\,d\phi_1d\phi_2\,\,, 
\end{equation}
\noindent
as in Eq.(18) or in the original paper of Born and v.Karman [9], Eq.(50) 
therein. The function f($\phi_1,\phi_2$) in Eq.(37) describing the frequency 
spectrum of the oscillations are the same for the diatomic neutrino pairs 
in the $\pi^\pm$ mesons and the diatomic neutrino pairs in the $\mu$ 
mesons because they must obey the group velocity limitation. When both 
functions are the same for the $\pi^\pm$ and the $\mu^\pm$ mesons the 
ratio of the energy in the diatomic $\bar{\nu}_\mu$-$\nu_e$ lattice 
oscillations of the $\mu$ mesons 
to the energy of the diatomic lattice oscillations in the $\pi^\pm$ mesons 
is 1/2 because the number of the pairs 
$\bar{\nu}_\mu$-$\nu_e$ in the $\mu$ mesons is 1/2 of the combined 
number N of the pairs 
$\bar{\nu}_\mu$-$\nu_e$ and $\nu_\mu$-$\bar{\nu}_e$ in the $\pi^\pm$ 
mesons. But since the same 
applies for the diatomic oscillations of the $\bar{\nu}_\mu$-$\bar{\nu}_e$ 
pairs in the $\mu^\pm$ mesons, the sum of the energies of both 
oscillations is equal to the oscillation energy of the $\pi^\pm$ mesons, 
as (b) says. 

   Finally we ask why do the $\mu$ mesons not interact strongly with the 
mesons and baryons? We will show in the next section that a strong force 
emanates from the sides of a cubic lattice caused by the unsaturated weak 
forces of about $10^6$ neutrinos at the surface of the neutrino lattice of 
the mesons and baryons. This follows from the study of Born and Stern [32] 
which dealt with the forces between two parts of a cubic lattice cleaved 
in vacuum. If the $\mu$ mesons have a lattice consisting of one type of 
muon neutrinos, say, 
$\bar{\nu}_\mu$ and of $\nu_e$ and $\bar{\nu}_e$ 
neutrinos their lattice surface is not the same as the surface of the cubic 
lattice of the mesons and baryons as described by the standing wave model. 
Therefore it does not seem likely that the $\mu$ mesons interact in the 
same way with the mesons and baryons as the mesons and baryons interact 
with each other. To put this in another way, the $\mu$ meson lattice does not 
bond with the cubic lattice of the mesons and baryons.

   To summarize: It has been shown that the mass of the $\mu^\pm$ mesons can 
be explained as the sum of the rest masses of 0.7$\cdot10^9$  muon 
neutrinos, respectively anti-muon neutrinos, as well as the same number of 
electron neutrinos and anti-electron neutrinos, plus their oscillation 
energies. The three neutrino types in the $\mu$ mesons are the remains of 
the cubic neutrino lattice of the $\pi^\pm$ mesons from which the $\mu$ 
mesons are formed in the $\pi^\pm$ decay. The energy contained in the sum of 
the rest masses of all muon or anti-muon neutrinos in the $\pi^\pm$ mesons 
is lost during the $\pi^\pm$ decay, whereas the oscillation energy of all 
neutrinos in the $\pi^\pm$ is preserved during the decay. Hence the mass of 
the $\mu^\pm$ mesons differs from the mass of the $\pi^\pm$ mesons by the 
sum of the energies of the rest masses of N/4 muon neutrinos which is 33.9 MeV $\cong$ 
1/4$\cdot$m($\pi^\pm)$c$_\star^2$ or its mass is m($\mu^\pm$) $\cong$ 
$3/4\cdot$m($\pi^\pm$). We have also found that the $\mu$ mesons do not interact with the mesons and baryons in the same 
strong way as the mesons and baryons interact with each other.

\section{The weak and strong nuclear force}

The potential of the force which holds a cubic ionic lattice together has 
been determined by Born and Land\'{e} (B\&L) [33]. They assumed that the 
static electric potential $\Phi$ of an elementary cube in the lattice has an 
attractive and repulsive part and is of the form

\begin{equation} \Phi = - \frac{\mathrm{a}}{\delta} + \frac{\mathrm{b}}{\delta^n}\,,
\end{equation}
\noindent
where $\delta$ is the distance between two ions of the same type along the 
edge of the crystal. The constant b is eliminated with the condition that 
$\Phi$ is a minimum in equilibrium. It follows that b = 
a$\delta_0^{n-1}$/n, where $\delta_0$ is the equilibrium distance between 
the two ions. The interaction constant a follows from the sum of the 
contributions of all ions of the lattice, using a method originally 
introduced by Madelung. According to Eq.(5) of B\&L it is then 

\begin{equation} \mathrm{a} = 13.94e^2\,,\end{equation}
\noindent
with the electron charge e. From (37) we arrive at 

\begin{equation} \Phi = \frac{\mathrm{a}}{\delta}\cdot(\frac{1}{n}(\frac{\delta_0}{\delta})^{n-1} - 
1)\,.\end{equation}
\noindent
The exponent n was determined by B\&L with the help of the compression 
modulus $\kappa$ given by Eq.(4) of B\&L

\begin{equation} \kappa = 9\delta_0^4/\mathrm{a}(n-1)\,.
\end{equation}
\noindent
In mechanics the value of the compression modulus is known. We follow the 
same route as B\&L and use, as we have done before in [34], for $\kappa$ 
the value which follows from the compression modulus of the nucleon 
K$_A$ = 900 to 1200 MeV, as determined theoretically by Bhaduri, Dey and Preston 
[35]. The value of K$_A$ found by them is supported by other theoretical 
and experimental studies of K$_{NM}$ of nuclei. Setting K$_A$ = 1000 MeV and 
using the formula K$_A$ = 9/$\rho\kappa$ from [35], where $\rho$ is the 
number density of nucleons per fm$^3$, we arrive with r$_p$ = 0.88\,fm at 
\begin{equation}\kappa = 1.603\cdot 10^{-35} \,\mathrm{cm^2/dyn}\,.
\end{equation}
\noindent
From Eq.(41) follows n\,$-$\,1 after we have replaced the 
electrostatic interaction e$^2$ in Eq.(39) with the weak 
interaction g$^2$, because the lattice is held together by the 
weak force. According to Perkins [16] 
\begin{equation} g^2 = 1.02\cdot10^{-5}(M{_W}/M{_P})^2\cdot 
4\pi\hbar c\,\cong\,2.946\cdot10^{-17}\,\mathrm{erg\cdot cm}\,,
\end{equation}
\noindent
where M$_W$ is the mass of the W boson and M$_P$ is the mass of the proton. 
It then follows with $\delta_0$ = 2r$_0$, where r$_0$ is the range of the 
weak force r$_0$ = 10$^{-16}$ cm, that 
\begin{equation}n-1 = \varepsilon = 2.187\cdot10^{-12}\,,
\end{equation}
\noindent
which differs slightly from the value of n $-$ 1 given in [34] because we 
now use r$_p$ = 0.88$\cdot10^{-13}$ cm, which means that the nucleon number 
density has decreased.
The potential of the weak force which holds the lattice 
together is then, neglecting a term on the order of $\varepsilon^2$,
\begin{equation}\Phi = -\frac{13.94g^2}{\delta}\cdot\varepsilon[1 - 
ln(\frac{\delta_0}{\delta})]\,.\end{equation}
\noindent
A graph $\Phi$ versus $\delta$ is in [34].

   Born [36] has used lattice theory to explain the elasticity constant 
c$_{11}$ of cubic crystals which we need to calculate the rest mass of the 
muon neutrino from Eq.(29). In general c$_{11}$ must be proportional to the 
inverse of the compressibility $\kappa$ because of the relation 
\begin{equation} 3/\kappa = c_{11} +2c_{12}\,. \end{equation}
Considering the sum of the forces acting between the lattice points Born 
arrived at an equation for c$_{11}$ which we approximated in Eq.(17) of [34] 
by 
\begin{equation}c_{11} = \frac{0.538e^2(n-1)}{r_0^4}\,,\end{equation}
making use of the relation that n = 1 + $\varepsilon$ with $\varepsilon$ = 
O($10^{-12}$).
For a cubic nuclear lattice we replace e$^2$ by g$^2$, use for (n $-$ 1) 
the value in Eq.(44), and r$_0$ is equal to $10^{-16}$ cm. It follows that the 
nuclear elasticity constant c$_{11}$ is then
\begin{equation} c_{11} = 3.466\cdot10^{35}\, \mathrm{dyn/cm^2}\,.
\end{equation}
The uncertainty 
of the value of c$_{11}$ is largely due to the uncertainty of n $-$ 1 which 
is caused by the uncertainty of the value of the compression modulus 
$\kappa$. Although the mass of the muon neutrino calculated from Eq.(29) with this 
value of c$_{11}$ is about 50\% larger than the mass of the muon neutrino 
determined from the decay of the $\mu$ meson it is important to be able to 
determine m($\nu_\mu$) theoretically from lattice theory. 

   A side of a NaCl monocrystal cleaved in vacuum exerts a strong 
attractive short-range force on the other side of the cleaved crystal. So does the 
side of a cubic nuclear lattice exert an attractive force on the side of 
another cubic nuclear lattice. In the standing wave model the strong force is the sum of 
the weak forces originating from the lattice points at the sides of the 
lattice. From the potential of the weak force between the lattice points we can 
determine the force by which one side of a nuclear lattice attracts the 
side of another nuclear lattice. As follows from Eq.(45) the force between two 
cells of a cubic nuclear lattice separated by the distance $\delta$ is 
given by 
\begin{equation} \frac{d\Phi}{d\delta} = - 
\frac{13.94g^2}{\delta^2}\,\varepsilon\,ln(\frac{\delta_0}{\delta})\,.
\end{equation}
The two cells will certainly not bond when the distance between the two 
cells exceeds the range of the weak nuclear force, if not at a smaller 
distance. The range of the weak force is given by \emph{a} = 10$^{-16}$ cm, 
and $\delta_0$ is 2\emph{a}. So the force required to separate the surfaces 
of two cubic lattice cells is at least
\begin{equation}F_w = -\, \frac{13.94g^2}{\emph{a}^2}\,\varepsilon\,ln2 = 
-\, 9.66\,\frac{g^2\varepsilon}{\emph{a}^2}\,\,. \end{equation}
The force required to separate two sides of an entire cubic nuclear lattice 
or vice versa the force by which one side of a nuclear lattice attracts 
the side of another nuclear lattice is then at least
\begin{equation} F_s = 
-\, 4.8\cdot10^6\,\frac{g^2\varepsilon}{\emph{a}^2}\,\,\mathrm{dyn}\,,
\end{equation}
if there are 2$\cdot10^6$ lattice points or 0.5$\cdot10^6$ elementary 
cubic cell surfaces 
at the side of the lattices.

   That means that the ratio of the strong nuclear force to the weak 
nuclear force is on the order of $10^6$. An accurate value of the ratio of the 
strong and weak nuclear forces is apparently not known from the experiments 
other than that it is said that the ratio of the strong nuclear interaction constant 
$\alpha_s$ to the weak interaction constant $\alpha_w$ is $\alpha_s \approx 
10^6\alpha_w$.


\section*{Conclusions}


   It is very natural to assume that the mesons and baryons consist of the 
particles into which they decay. From the well-known decays follows that 
the particle spectrum consists of a $\gamma$-branch and a neutrino branch. 
 We also take for 
granted that the elementary integer multiple rule for the ratios of the 
masses of the mesons and baryons is a consequence of the structure of the 
particles. We take it for granted that Fourier analysis holds in high energy 
collisions of the particles, which means that a continuum of high 
frequencies must be present in the collision products. And we believe that the consensus of opinion that   
the weak nuclear force has a range on the order of 10$^{-16}$\,cm is realistic. On 
these points we base our model of the mesons and baryons.

   We have studied cubic nuclear lattices consisting of either photons or neutrinos 
whose lattice points are 10$^{-16}$\,cm apart. After a high energy collision 
the lattice points must oscillate and thereby satisfy the requirement 
that a continuous spectrum of high frequencies must be present in the 
mesons and baryons. The oscillations are in the form of standing waves. 
Standing waves do not have 
to propagate, so the particles can have a rest mass although the particles of 
the $\gamma$-branch consist of electromagnetic waves. The frequency 
spectrum of the waves is determined by the group velocity which cannot 
exceed the velocity of light. The sum of the energies contained in the 
frequencies of all standing waves determines the mass of the particles. 
Because of the group velocity limitation the energy of the different 
oscillation modes or superpositions of modes are integer multiples of the 
basic mode. The masses of the particles of the $\gamma$-branch are 
therefore integer multiples of the basic mode or of the $\pi^0$ meson, in 
agreement with what the ratios of the particle masses strongly suggest. 
Each particle has automatically an antiparticle. The size of the particles 
is limited by the radiation pressure to about 10$^{-13}$ cm. The absolute 
value of the energy of the $\pi^0$ meson is that of a cubic black body 
filled with standing electromagnetic waves whose energy is determined by 
Planck's classical formula for the energy of a linear oscillator. \emph{The 
masses of the particles of the $\gamma$-branch can thus be explained using 
photons only}. A very conservative explanation which avoids the introduction 
of any new particle.

   As the decay of the neutrino branch particles suggests we assume that 
they consist of muon and electron neutrinos and their antiparticles. The 
oscillations of the neutrino lattice contain the continuum of frequencies 
which must be in the $\nu$-branch particles after their creation. The 
energy in the $\nu$-branch particles is not only the energy of 
all lattice oscillations but is the sum of the energy in 
the neutrino lattice oscillations and the energy in the rest masses of 
the $10^9$ neutrinos making up the lattice. \emph{The $\pi^\pm$ mesons consist of the 
four types of neutrinos and their oscillation energies, plus electric charge.} Their mass so 
determined is 0.98m($\pi^\pm$)(exp). On the other hand, a $\pi^0$ meson has 
to be superposed on the (2.2) mode of the $\pi^\pm$ meson oscillations in order to 
explain the mass of the K$^\pm$ mesons. Then it is natural that $\pi^0$ 
mesons appear in 29\% of the K$^\pm$ decays. For the explanation of the 
higher modes of the particles of the $\nu$-branch it is necessary to 
consider \emph{neutrinos and standing electromagnetic waves} and charge.

   The mass of the $\mu$ mesons can also be explained with a neutrino 
lattice. The $\mu$ mesons consist of the neutrinos remaining after the 
decay of the $\pi^\pm$ neutrino lattice. All muon neutrinos of one type, 
either the muon neutrinos or the anti-muon neutrinos, are removed from the 
neutrino lattice of the $\pi^\pm$ mesons in its decay. That releases 
the sum of the energy of their rest masses which is 
$\cong$\,1/4$\cdot$m($\pi^\pm$)c$_\star^2$. Since it can be shown that the 
oscillation energy of all neutrinos in the $\pi^\pm$ mesons is conserved in 
the decay, the energy of the rest mass of the $\mu^\pm$ mesons is 
$\cong$\,3/4$\cdot$m($\pi^\pm$)c$_\star^2$, in near agreement with the exact ratio 
m($\mu^\pm$)/m($\pi^\pm$) = 0.757028.  From 
the decay of the $\pi^\pm$ mesons follows that \emph{the rest mass of the 
muon neutrino is} m($\nu_\mu)$ = 47.5 milli-eV/c$_\star^2$. From the decay of the neutron 
follows that the \emph{rest mass of the electron neutrino is} m($\nu_e)$ = 0.55 meV/c$_\star^2$.

   Born's lattice theory provides the means to determine the potential of 
the force which extends from one lattice point to the next and thereby 
holds the lattice together. Following Born's approach exactly, replacing 
only e$^2$ by the weak interaction constant g$^2$, we learn that at the 
lattice distance \emph{a} the repulsive term of the potential differs from the 
attractive term by only 10$^{-12}$. We also learn that from the weak force 
in the lattice follows automatically the existence of the strong nuclear force 
emanating from the sides of the lattice, caused by the 10$^6$ unsaturated 
weak forces at a side of the lattice.

   We conclude that the standing wave model solves a number of problems 
for which an answer heretofore has been hard to come by. Only photons and 
neutrinos and charge are needed to explain the stable mesons and baryons. 
In a forthcoming paper we will show that the spin of the baryons can also 
be explained with the standing wave model.

\bigskip

   {\bfseries Acknowledgment}. I gratefully acknowledge the contributions of
 
Dr. T. Koschmieder to this study.

\section*{References}

\noindent
[1] Gell-Mann,M. Phys.Lett.B. {\bfseries111},1 (1964).

\smallskip
\noindent
[2] Barnett,R. et al. Rev.Mod.Phys. {\bfseries68},611 (1996).

\smallskip
\noindent
[3] Witten,W. Physics Today {\bfseries X},24 (1996).

\smallskip
\noindent
[4] Koschmieder,E.L. Bull.Acad.Roy.Belgique {\bfseries X},281 (1999).\\
\indent  arXiv: hep-ph/0002179(2000).

\smallskip
\noindent
[5] Koschmieder,E.L. Nuovo Cim. {\bfseries99},555 (1988).

\smallskip
\noindent
[6] Koschmieder,E.L. and Koschmieder,T.H.\\
\indent Bull.Acad.Roy.Belgique {\bfseries X},289 (1999).\\
\indent arXiv: hep-lat/0002016 (2000).

\smallskip 
\noindent
[7] Wilson,K. Phys.Rev. {\bfseries D10},2445 (1974).

\smallskip
\noindent
[8] Weingarten,D. Scient.Am. {\bfseries274},116 (1996).

\smallskip
\noindent
[9] Born,M. and v.Karman,Th. Phys.Z. {\bfseries13},297 (1912).

\smallskip
\noindent
[10] Blackman,M. Proc.Roy.Soc. {\bfseries A148},365;384 (1935).

\smallskip
\noindent
[11] Blackman,M. Handbuch der Physik VII/1, Sec.12 (1955).

\smallskip
\noindent
[12] Born,M. and Huang,K. \emph{Dynamical Theory of Crystal Lattices},\\ 
\indent \,\,(Oxford) (1954).

\smallskip
\noindent
[13] Maradudin,A. et al. \emph{Theory of Lattice Dynamics in the Harmonic\\ 
\indent \,\, Approximation}, Academic Press, 2nd edition, (1971).

\smallskip
\noindent
[14] Ghatak,A.K. and Khotari,L.S. \emph{An introduction to Lattice Dynamics},\\
\indent \,\, Addison-Wesley, (1972).

\smallskip
\noindent
[15] Schwinger,J. Phys.Rev. {\bfseries128},2425 (1962).

\smallskip
\noindent
[16] Perkins,D.H. \emph{Introduction to High-Energy Physics},\\
\indent \,\,Addison Wesley, (1982).

\smallskip
\noindent
[17] Rosenfelder,R. arXiv: nucl-th/9912031 (2000).

\smallskip
\noindent
[18] Born,M. Proc.Camb.Phil.Soc. {\bfseries36},160 (1940).

\smallskip
\noindent
[19] Koschmieder,E.L. arXiv: hep-lat/0005027 (2000).

\smallskip
\noindent
[20] Liesenfeld,A. et al. Phys.Lett.B {\bfseries468},20 (1999).

\smallskip
\noindent
[21] Bernard,V., Kaiser,N. and Meissner,U-G.\\
\indent  \,\,arXiv: nucl-th/0003062 
(2000).

\smallskip
\noindent
[22] Sommerfeld,A. \emph{Vorlesungen \"{u}ber Theoretische Physik},\\
\indent  \,\,Bd.V, p.56 (1952).

\smallskip
\noindent
[23] Debye,P. Ann. d. Phys. {\bfseries39},789 (1912).

\smallskip
\noindent
[24] Bose,S. Zeitschr. f. Phys. {\bfseries26},178 (1924).

\smallskip
\noindent
[25] Koschmieder,E.L. and Koschmieder,T.H.\\
\indent  \,\,arXiv: hep-lat/0104016 (2001).

\smallskip
\noindent
[26] Thirring,H. Phys.Z. {\bfseries15},127. (1914).

\smallskip
\noindent
[27] Bethe,H. Phys.Rev.Lett. {\bfseries58},2722 (1986).

\smallskip
\noindent
[28] Bahcall,J.N. Rev.Mod.Phys. {\bfseries59},505 (1987).

\smallskip
\noindent
[29] Fukuda,Y. et al. Phys.Rev.Lett. {\bfseries81},1562 (1998).

\smallskip
\noindent
[30] Ahmad,Q.R. et al. Phys.Rev.Lett. {\bfseries87},071301 (2001).

\smallskip
\noindent
[31] Koschmieder,E.L. arXiv: physics/0110005 (2001).

\smallskip
\noindent
[32] Born,M. and Stern,O. Sitzungsber.Preuss.Akad.Wiss.\\ 
\indent  \,\,{\bfseries33},901 (1919).

\smallskip
\noindent
[33] Born,M. and Land\'{e},A. Verh.Dtsch.Phys.Ges.\\
\indent \,\,{\bfseries20},210 (1918).

\smallskip
\noindent
[34] Koschmieder,E.L. Nuovo Cim. {\bfseries101},1017 (1989).

\smallskip
\noindent
[35] Bhaduri,R.K., Dey,J. and Preston,M.A. Phys.Lett.B {\bfseries136},\\
\indent \,\,289 (1984).

\smallskip
\noindent
[36] Born,M. Ann.Phys. {\bfseries61},87 (1920).

\end{document}